\documentclass{aa}  

\usepackage{natbib}
\bibpunct{(}{)}{;}{a}{}{,}
\usepackage{graphicx}
\usepackage{txfonts}
\usepackage{url}
\usepackage{orcidlink}
\hypersetup{colorlinks=true,linkcolor=black,citecolor=blue,filecolor=blue,urlcolor=blue}
\RequirePackage{silence}
\begin{document} 

   \title{Enriched volatiles and refractories but deficient titanium on the dayside atmosphere of WASP-121b revealed by JWST/NIRISS}

   \author{
            S. Pelletier\inst{1, 2}\thanks{Corresponding author: \texttt{Stefan.Pelletier@unige.ch}} \orcidlink{0000-0002-8573-805X},
            L.-P. Coulombe\inst{2} \orcidlink{0000-0002-2195-735X},
            J. Splinter\inst{3,4} \orcidlink{0000-0001-9987-467X},
            B. Benneke\inst{5,2} \orcidlink{0000-0001-5578-1498},
            R.~J. MacDonald\inst{6,7} \orcidlink{0000-0003-4816-3469},
            D. Lafreni\`ere\inst{2} \orcidlink{0000-0002-6780-4252},
            N.~B. Cowan\inst{8,4} \orcidlink{0000-0001-6129-5699},
            R. Allart\inst{2}\thanks{SNSF Postdoctoral Fellow} \orcidlink{0000-0002-1199-9759},
            E. Rauscher\inst{6} \orcidlink{0000-0003-3963-9672},
            R.~C. Frazier\inst{6} \orcidlink{0000-0001-6569-3731},
            M.~R. Meyer\inst{6} \orcidlink{0000-0003-1227-3084},
            L. Albert\inst{2,9} \orcidlink{0000-0003-0475-9375},
            L. Dang\inst{10,2} \orcidlink{0000-0003-4987-6591},
            R. Doyon\inst{2,9} \orcidlink{0000-0001-5485-4675},
            D. Ehrenreich\inst{1} \orcidlink{0000-0001-9704-5405},
            L. Flagg\inst{11,12} \orcidlink{0000-0001-6362-0571},
            D. Johnstone\inst{13,14} \orcidlink{0000-0002-6773-459X},
            A.~B. Langeveld\inst{15,16} \orcidlink{0000-0002-4451-1705},
            O. Lim\inst{2} \orcidlink{0000-0003-4676-0622},
            C. Piaulet-Ghorayeb\inst{17}\thanks{E.~Margaret Burbridge Postdoctoral Fellow} \orcidlink{0000-0002-2875-917X},
            M. Radica\inst{17}\thanks{NSERC Postdoctoral Fellow} \orcidlink{0000-0002-3328-1203},
            J. Rowe\inst{18} \orcidlink{0000-0002-5904-1865},
            J. Taylor\inst{19} \orcidlink{0000-0003-4844-9838},
            J.~D. Turner\inst{20} \orcidlink{0000-0001-7836-1787}
           }

   \institute{
            Observatoire astronomique de l'Universit\'e de Gen\`eve, 51 chemin Pegasi 1290 Versoix, Switzerland
        \and
            Trottier Institute for Research on Exoplanets, Department of Physics, Universit\'e de Montr\'eal, 1375 Avenue Th\'er\`ese-Lavoie-Roux, Montr\'eal, QC H2V 0B3, Canada
        \and
            Trottier Space Institute at McGill, 3550 rue University, Montr\'eal, QC H3A 2A7, Canada
        \and
            Department of Earth \& Planetary Sciences, McGill University, 3450 rue University, Montreal, QC H3A OE8, Canada
        \and
            Department of Earth, Planetary, and Space Sciences, University of California, Los Angeles, CA, USA
        \and
            Department of Astronomy, University of Michigan, 1085 S. University Ave., Ann Arbor, MI 48109, USA
        \and
            School of Physics and Astronomy, University of St Andrews, North Haugh, St Andrews, KY16 9SS, UK
        \and
            Department of Physics, McGill University, 3600 rue University, Montr\'eal, QC H3A 2T8, Canada
        \and
            Observatoire du Mont-M\'egantic, Universit\'e de Montr\'eal, C.P. 6128, Succ. Centre-ville, Montr\'eal, QC H3C 3J7, Canada
        \and
            Department of Physics and Astronomy, University of Waterloo, 200 University W, Waterloo, ON N2L 3G1, Canada
        \and
            Department of Astronomy \& Astrophysics, 525 Davey Laboratory, The Pennsylvania State University, University Park, PA, 16802, USA
        \and
            Center for Exoplanets and Habitable Worlds, 525 Davey Laboratory, The Pennsylvania State University, University Park, PA, 16802, USA 
        \and
            NRC Herzberg Astronomy and Astrophysics, 5071 West Saanich Rd, Victoria, BC, V9E 2E7, Canada 
        \and
            Department of Physics and Astronomy, University of Victoria, Victoria, BC V8P 5C2, Canada
        \and
            Department of Physics and Astronomy, Johns Hopkins University, Baltimore, MD 21218, USA
        \and
            Department of Astronomy and Carl Sagan Institute, Cornell University, Ithaca, NY 14850, USA
        \and
            Department of Astronomy \& Astrophysics, University of Chicago, 5640 South Ellis Avenue, Chicago, IL 60637, USA
        \and
            Department of Physics \& Astronomy, Bishop's University, 2600 Rue College, Sherbrooke, QC J1M 1Z7, Canada
        \and
            Department of Physics, University of Oxford, Parks Rd, Oxford, OX1 3PU, UK
        \and
            Department of Astronomy and Carl Sagan Institute, Cornell University, Ithaca, NY 14853, USA
             }

   \date{Received XXX; accepted XXX}

  \abstract
   {
   With dayside temperatures elevated enough for all atmospheric constituents to be present in gas form, ultra-hot Jupiters offer a unique opportunity to probe the composition of giant planets.
    }
   {
   We aim to infer the composition and thermal structure of the dayside atmosphere of the ultra-hot Jupiter WASP-121b from two NIRISS$/$SOSS secondary eclipses observed as part of a full phase curve.
   }
   {
   We extract the eclipse spectrum of WASP-121b with two independent data reduction pipelines and analyse it using different atmospheric retrieval prescriptions to explore the effects of thermal dissociation, reflected light, and titanium condensation on the inferred atmospheric properties.
   }
   {
    We find that the observed dayside spectrum of WASP-121b is best fit by atmosphere models possessing a stratospheric inversion with temperatures reaching over 3000\,K, with spectral contributions from H$_2$O, CO, VO, H$^{-}$, and either TiO or reflected light. We measure the atmosphere of WASP-121b to be metal enriched ($\sim$10$\times$ stellar) but comparatively titanium poor ($\lesssim$1$\times$ stellar), potentially due to partial cold-trapping. The inferred C$/$O depends on model assumptions such as whether reflected light is included, ranging from being consistent with stellar if a geometric albedo of zero is assumed to being super-stellar for a freely fitted $A_g = 0.16_{-0.02}^{+0.02}$. The volatile-to-refractory ratio is measured to be consistent with the stellar value.
   }
   {
   From the NIRISS eclipse spectrum, we infer that WASP-121b has an atmosphere enriched in both volatile and refractory metals, but not in ultra-refractory titanium, suggesting the presence of a nightside cold-trap. Considering H$_2$O dissociation is critical in free retrieval analyses, leading to order-of-magnitude differences in retrieved abundances for WASP-121b if neglected. Simple chemical equilibrium retrievals assuming that all species are governed by a single metallicity parameter drastically overpredict the TiO abundance, strongly biasing the inferred atmospheric composition.
   }

   \keywords{planets and satellites: gaseous planets --
                planets and satellites: atmospheres --
                planets and satellites: composition --
                planets and satellites: formation --
                planets and satellites: individual: WASP-121b --
                techniques: spectroscopic
               }

    \authorrunning{Pelletier et al.}
    \titlerunning{NIRISS eclipse of WASP-121b}

   \maketitle

\section{Introduction}

With temperatures exceeding $\sim$2200\,K, ultra-hot Jupiter daysides are hot enough to vapourise even the most refractory (high condensation temperature) compounds~\citep{lodders_solar_2003}. As a result, their atmospheres are expected to be mostly cloud-free with the full chemical inventory of species in the gas phase accessible to observations~\citep{helling_cloud_2021}. This stands in contrast to the vast majority of currently known giant exoplanets which are in temperature regimes where only volatiles (e.g., C, O) and moderately refractory species (e.g., S) can exist in gaseous form while rock-forming refractory elements (e.g., Fe, Mg, Si) are condensed out of the upper atmosphere, inaccessible to remote sensing~\citep{lewis_clouds_1969}. Ultra-hot Jupiters thus provide an unprecedented opportunity to probe the unsequestered bulk volatile and refractory composition of giant planetary envelopes, measurements that remain elusive even for the Solar System giants~\citep[e.g.,][]{mousis_determination_2009, mousis_key_2020}.

The bulk composition of a giant planet's primordial envelope is the result of its accretion history, of which the present-day atmosphere may still hold an imprint of~\citep[e.g.,][]{mordasini_imprint_2016}. As such, constraining the relative proportions of volatiles and refractory (a proxy for the ice-to-rock ratio) is of particularly high interest from a planet formation and evolution standpoint as it can allow for degeneracies regarding how material was accreted to be broken~\citep{turrini_tracing_2021, lothringer_new_2021, pacetti_chemical_2022, schneider_how_2021, chachan_breaking_2023, crossfield_volatile--sulfur_2023}. For example, while H$_2$O can be accreted as both vapour and ice depending on the local temperature, refractories will almost always be accreted as solids.  Measurements of refractory elements in a planet's atmosphere can therefore provide an absolute measure of how much rocky material was accreted during formation~\citep{lothringer_new_2021}.

Unlike for colder planets where refractory compounds are condensed out of the gas phase, the vapourisation of refractories in ultra-hot Jupiter dayside atmospheres can provide a wealth of additional opacity sources to the atmosphere. Notably, refractory atomic metals and their oxides have preferentially large cross sections at optical wavelengths compared to volatile molecules (e.g., H$_2$O, CO) that tend to have stronger bands in the infrared~\citep[e.g.,][]{gandhi_molecular_2020, pelletier_crires_2025}. In combination with intense stellar irradiation, strong optical absorbers such as TiO can lead to stellar energy being deposited at lower pressures, adding heat to the upper atmosphere and forming a stratosphere~\citep{hubeny_possible_2003, fortney_unified_2008}. The presence of hot stratosphere can then further drive the dissociation of molecules and ionization of metals, changing the chemistry of the atmosphere.
For eclipse observations, a change from a negative to a positive lapse rate at photospheric pressures will also change the thermal spectrum of an exoplanet from showing absorption, to showing emission lines~\citep[e.g.,][]{haynes_spectroscopic_2015}.

The ultra-hot Jupiter WASP-121b is one of the most extensively studied exoplanets to date. And yet, investigations of its atmosphere have left us with more questions than answers. In particular, WASP-121b has been at the forefront of numerous open ended questions in the field. Are stratospheres in ultra-hot Jupiters driven by TiO as first predicted~\citep{hubeny_possible_2003, fortney_unified_2008}, or rather by a combination of other optical absorbers~\citep{spiegel_can_2009, lothringer_extremely_2018, gandhi_new_2019, piette_assessing_2020}? Does the nightside of WASP-121b have a cold-trap depleting titanium species from the gas phase in both its dayside and terminator~\citep{evans_optical_2018, mikal-evans_diurnal_2022, hoeijmakers_hot_2020, hoeijmakers_mantis_2024}? How relevant is H$_2$O dissociation in shaping its spectrum~\citep{parmentier_thermal_2018, bazinet_quantifying_2025}? Can a giant planet's accretion history leave an imprint on its atmospheric volatile-to-refractory ratio~\citep{lothringer_new_2021, chachan_breaking_2023, smith_roasting_2024, pelletier_crires_2025, evans-soma_sio_2025}?

In this paper we analyse new observations of WASP-121b taken with NIRISS$/$SOSS onboard the JWST in the context of these points of intrigue. Section~\ref{sec:data_reduction} detais the data and how it was extracted to produce a spectrum.  The atmospheric models used to analyse said spectrum are described in Section~\ref{sec:modelling}. We present and discuss our results in Section~\ref{sec:results} and conclude in Section~\ref{sec:conclusion}.

\section{Observations and Data Extraction} \label{sec:data_reduction}
We observed a full phase curve of the ultra-hot Jupiter WASP-121b ($R_p = 1.753\pm0.036$\,$R_{\rm Jup}$, $M_p = 1.157\pm0.070$\,$M_{\rm Jup}$, $P = 1.2749250(1)$\,days)~\citep{delrez_wasp-121_2016, bourrier_optical_2020, patel_empirical_2022} using the SOSS~\citep{albert_near_2023} mode of the NIRISS instrument~\citep{doyon_near_2023} on board the JWST. The data were taken as part of the NIRISS Exploration of the Atmospheric diversity of Transiting exoplanets (NEAT) Guaranteed Time Observation Program (GTO 1201; PI D.~Lafreni\`ere).  The observations were taken starting UT 17:39:55 26 October 2023 and consist of 3452 integrations (6 groups each) using the SUBSTRIP256 subarray ($\lambda=0.6-2.83\,\mu$m).  The full time sequence lasted a total of 36.9 hours, beginning before secondary eclipse, covering a full orbital period of WASP-121b, and then ending after a second secondary eclipse.  In this work we focus on characterising the dayside atmosphere of WASP-121b via the extracted spectrum from the combined secondary eclipses.  We refer the reader to companion analyses for the transit (MacDonald et al.~in prep.) and phase curve \citep{splinter_precise_2025, frazier_days_2025}. The NIRISS data used in this work can be found here\footnote{\url{https://doi.org/10.17909/gg5q-ct70}}.

\subsection{Data Reduction}\label{subsec:data_reduction}

\begin{figure}
    \centering
    \includegraphics[width=0.49\textwidth]{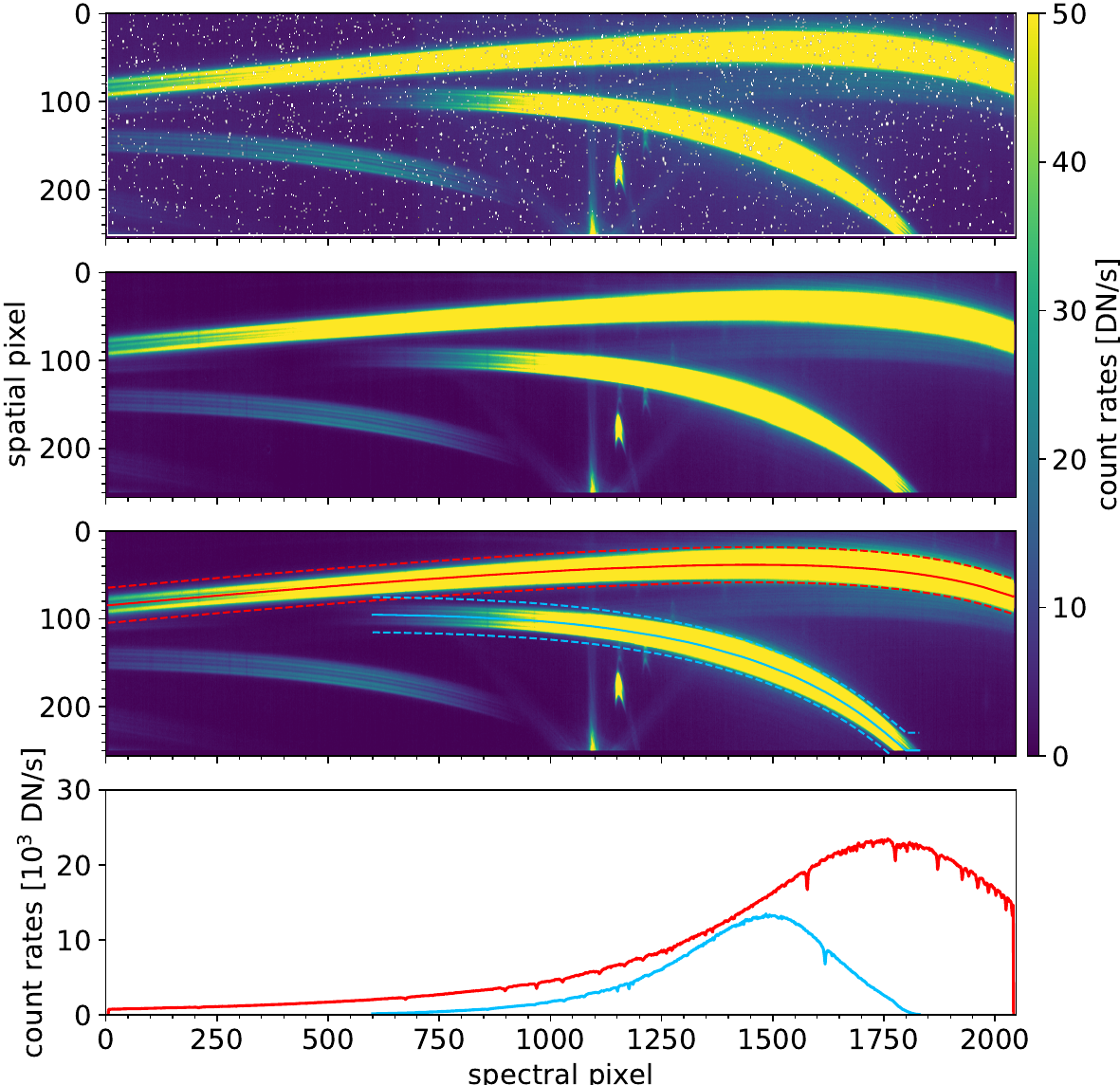}
    \vspace{-3mm}
    \caption{Reduction steps and spectral extraction of the NIRISS/SOSS data.
    Top panel: NIRISS detector after subtraction of the super-bias and non-linearity correction, and ramp-fitting, converted from counts to count rates. Second panel: After bad pixel and cosmic ray hits correction, and subtraction of the non-uniform background. Third panel: Frame after correction of the $1/f$ noise. Bottom panel: Extracted spectra for the first (red) and second (blue) spectral orders.
    }
    \label{fig:reduction_steps}
    \vspace{-3mm}
\end{figure}

We reduce the NIRISS$/$SOSS time series observations of the phase curve of WASP-121\,b using the \texttt{NAMELESS} pipeline~\citep{coulombe_broadband_2023, coulombe_highly_2025}. We start from the raw uncalibrated data and go through all Stage 1 and Stage 2 steps of the STScI \texttt{jwst} pipeline v1.12.5 \citep{bushouse_jwst_2023} to perform superbias subtraction, reference pixel correction, linearity correction, jump detection, ramp fitting, and flatfield correction. We then proceed with a series of custom reduction steps to correct for bad pixels, non-uniform background, cosmic rays, and $1/f$ noise (Fig.~\ref{fig:reduction_steps}) following the same methodology outlined in \cite{allart_complex_2025}. We correct for bad pixels by dividing the detector into a series of 4$\times$4 pixel windows for each integration, in which each pixel whose spatial second derivative is more than 4$\sigma$ away from the median of its window is flagged as bad. Any pixel that is flagged in more than 50\% of all integrations is then considered a bad pixel and is interpolated over all integrations using the bicubic interpolation function \texttt{scipy.interpolate.griddata}.

To deal with the non-uniform background, we follow the methodology of \citet{lim_atmospheric_2023} and split the STScI model background into two distinct sections separated by the jump in background flux that occurs around spectral pixel x$\sim$700. We then scale each section separately using portions ($x_1\in[500,650]$, $y_1\in[230,250]$) and ($x_2\in[740,850]$, $y_2\in[230,250]$) of the detector and the 16$^\mathrm{th}$ percentile of the ratio of the median frame as the scaling factor to the model background. 

We then look for any remaining cosmic rays not flagged by the \texttt{jwst} pipeline jump detection step. To avoid flagging outliers caused by $1/f$ noise as cosmic rays, we first perform a temporary correction of the $1/f$ noise by subtracting the median frame scaled by the white-light curve (obtained by summing the counts from pixels $x\in[1200,1800]$, $y\in[25,55]$) from all integrations, and then subtracting the median from all columns. After this preliminary $1/f$ correction, we then compute the running median in time of all individual pixels and flag any pixel deviating by more than 4$\sigma$ at a given integration. These counts are subsequently set to the running median value and the previously removed $1/f$ noise is added back to be corrected at a later stage.

We measure the central position of the spectral orders at each column by cross-correlating the detector columns of the median frame with a template trace profile. For the template trace profile, we use column x = 2038 of the order 1 trace model contained in reference file \texttt{jwst\_niriss\_specprofile\_0022.fits}\footnote{\url{https://jwst-crds.stsci.edu/browse/jwst_niriss_specprofile_0022.fits}}. We then cross-correlate this template with all detector columns, where we super-sample both the template and the data by a factor of 1000 and take the maximum cross-correlation function position as the trace centre for a given column. This process is repeated twice for order 1 and 2.

We correct for the $1/f$ noise using the same methodology presented in \cite{coulombe_highly_2025}, applying a different scaling value for each column and each trace, which is especially important for a phase curve since the planetary flux is likely to show a strong wavelength dependence \citep{coulombe_broadband_2023}. After having performed all corrections on the detector images, we then extract the flux from the first and second spectral orders using a box aperture. We extract the flux for box widths of 30, 36, and 40 pixels and find consistent (within $\pm$2\%) values of the point-to-point scatter for all wavelengths. We thus proceed with a box width of 40 pixels to minimise our sensitivity to variations in the trace morphology \citep{coulombe_broadband_2023}. We identify pixel regions overlapping with background sources: pixels 1075 to 1105 (1.7717 to 1.7423\,$\mu$m) in Order 1 as well as pixels 1300 to 1425 (0.7977 to 0.7410\,$\mu$m) and $<$1230 ($>$0.8299\,$\mu$m) in Order 2. We later verify that our results are not affected by any of these potentially contaminated wavelengths.

We perform our wavelength calibration using the \texttt{PASTASOSS} python package \citep{baines_characterization_2023, baines_characterization_2023-1} which takes as input the pupil wheel position for a given time series observation ($\theta_\mathrm{pupil}$ = 245.766$^\circ$ in this case) and returns the position of the spectral traces as well as the wavelength solutions for both the first and second spectral orders.

\subsection{Light Curve Fitting}

For the light curve fitting, we first construct a model that accounts for the primary transit, secondary eclipses, phase-curve modulation, and systematics. For the astrophysical model, we define the system flux, normalised by the mean stellar flux, as the sum of the transit $\mathcal{T}$ and eclipse $\mathcal{E}$ functions, the latter of which is multiplied by planet-to-star-flux ratio

\begin{equation}\label{eq:system_flux}
f(t) = \mathcal{T}(t,\Omega)\frac{F_\mathrm{s}(t)}{\bar{F_\mathrm{s}}} + \mathcal{E}(t,\Omega) \frac{F_\mathrm{p}(t)}{F_\mathrm{s}(t)}. 
\end{equation}

\noindent 
The shape of the transit and eclipse functions depends on the system parameters ($\Omega$), which include the mid-transit time ($T_0$), orbital period ($P$), planet-to-star radius ratio ($R_\mathrm{p}/R_\mathrm{s}$), semi-major axis ($a/R_{s}$), impact parameter ($b$), eccentricity ($e$), and argument of periapsis ($\omega$). The transit shape also depends on the intensity profile of the star, which we model using the quadratic limb-darkening law with coefficients [$u_1,u_2$]. The transit function is multiplied by the mean-normalised stellar flux considering the Doppler boosting and ellipsoidal variation effects ($F_\mathrm{s}(t)/\bar{F_\mathrm{s}} = 1 - D\sin\varphi - E\cos(2\varphi)$, \citealt{shporer_astrophysics_2017}). We use the \texttt{batman} \citep{kreidberg_batman_2015} python package to compute $\mathcal{T}(t)$ and $\mathcal{E}(t)$. For the phase-dependent planetary flux, we consider the form of \cite{cowan_inverting_2008} and express it as a second-order sinusoid 

\begin{equation}\label{eq:eclipse_depth_data}
\frac{F_p(t)}{\bar{F_s}} = \sum_{i=0}^{2} F_n \cos\left(n\left[\varphi(t) - \delta_n\right] \right) ,
\end{equation}

\noindent
where $n$ corresponds to the order of the sinusoid, $F_n$ is the amplitude of that order, and $\delta_n$ is the offset of the maximum of the sinusoid relative to the phase of mid-eclipse. The orbital phase is given by $\varphi = 2\pi (t-T_{\mathrm{sec}})/P$. The time of mid-eclipse is defined as $T_{\mathrm{sec}} = T_0 + \frac{P}{2} + \frac{2P}{\pi}e\cos\omega + \frac{2a}{c}$, such that we account for the delay in eclipse time for an eccentric orbit \citep[$\frac{2P}{\pi}e\cos\omega$;][]{winn_transits_2014}, as well as the light-time travel delay ($\frac{2a}{c} \approx $ 25.5 seconds).

\begin{figure}
    \centering
    \includegraphics[width=0.49\textwidth]{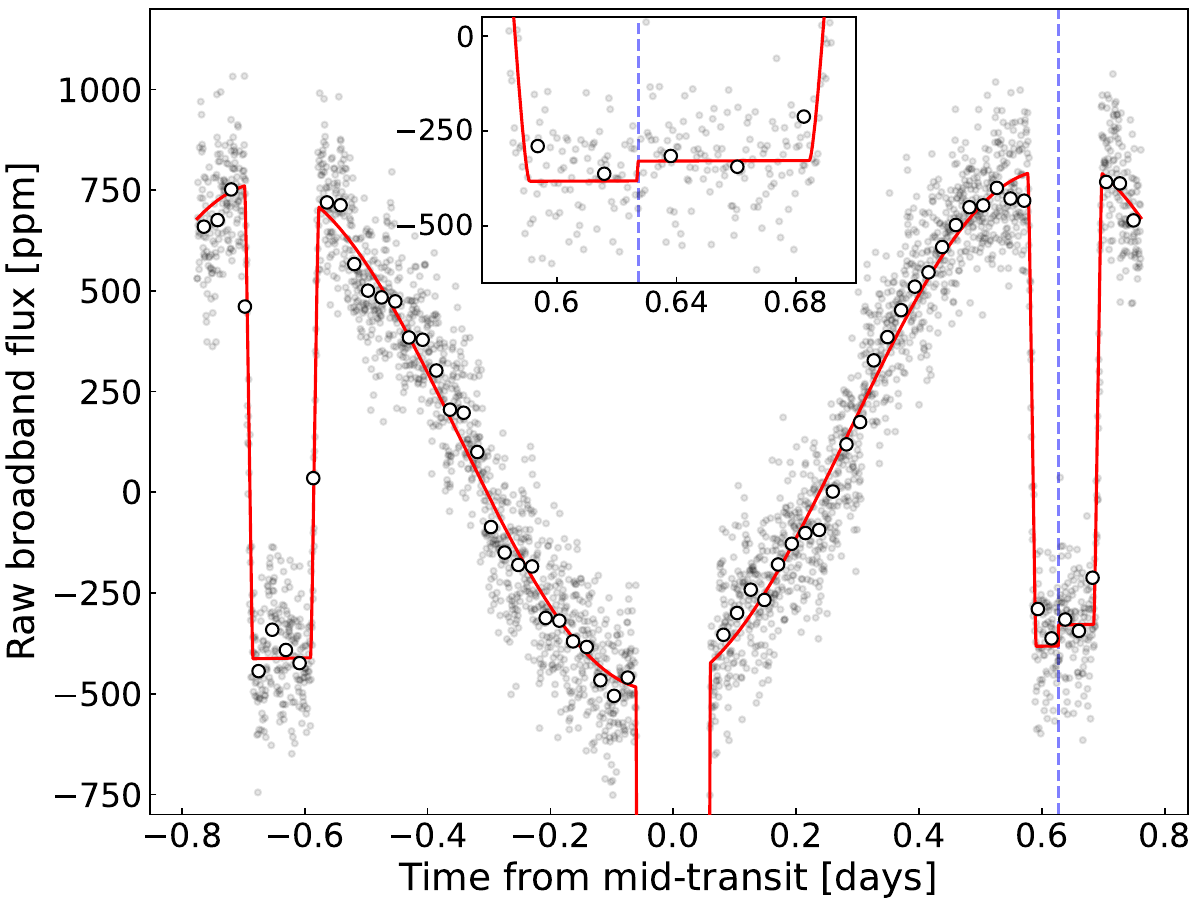}
    \vspace{-5mm}
    \caption{Observed NIRISS/SOSS raw white-light phase curve of WASP-121b and model fit. 
    The data are indicated as grey points (50-integration bins shown as white dots for clarity), with the combined astrophysical and systematics model shown in red. The tilt event occurring during the second eclipse (zoom in shown in the inset panel) is marked by the dashed blue line.
    }
    \label{fig:raw_wlc}
    \vspace{-3mm}
\end{figure}

For the systematics models, we assume the form
\begin{equation}\label{eq:systematic}
S(t) = c + v \cdot (t-t_0) + j \cdot \Theta(t-t_\mathrm{tilt}),
\end{equation}
and fit it following \cite{coulombe_highly_2025}, to which we refer the reader for more details. Here $\Theta$ is the Heaviside function to model a tilt event occurring during the second secondary eclipse~\citep[][see also Fig.~\ref{fig:raw_wlc}]{splinter_precise_2025}.

\subsubsection{White-Light Curve Fit}
We begin by fitting the white-light curve, constructed by summing all pixels of the first spectral order ($\lambda$ = 0.85--2.85\,$\mu$m), considering the astrophysical and systematics models of Eqs.~\ref{eq:system_flux} and \ref{eq:systematic}. We fit for the system parameters: $T_0$ ($\mathcal{U}[60244.45,60244.55]$\,BJD), $P$ ($\mathcal{N}[1.2749250,0.0000001^2]$, \citealt{patel_empirical_2022}), $R_\mathrm{p}/R_\mathrm{s}$ ($\mathcal{U}[0.01,0.2]$), $a/R_\mathrm{s}$ ($\mathcal{U}[1,20]$), $b$ ($\mathcal{U}[0,1]$), $u_{1,2}$ ($\mathcal{U}[-3,3]$), \citealt{coulombe_biases_2024}); the phase curve parameters: $F_{0}$ ($\mathcal{U}[-5000,5000]$\,ppm), $F_{1,2}$ ($\mathcal{U}[0,5000]$\,ppm), $\delta_{1,2}$ ($\mathcal{U}[-\pi/n,\pi/n]$); the systematics model parameters $c$ ($\mathcal{U}[-10^{9},10^{9}]$), $v$ ($\mathcal{U}[-10^5,10^5]$\,ppm$/$day), $j$ ($\mathcal{U}[-1000,1000]$\,ppm); and a photometric scatter value $\sigma$ ($\mathcal{U}[50,10^4]$\,ppm), for a total of 16 free parameters. For our nominal white-light curve fit, we assume a circular orbit ($e$ = 0) as well as null amplitudes for the Doppler boosting and ellipsoidal effects ($D,E$ = 0). Tests allowing for free eccentricity and Doppler/ellipsoidal amplitudes are described below. The parameter space exploration is done using the Markov chain Monte Carlo sampler \texttt{emcee}~\citep{foreman-mackey_emcee_2013}. The best fit model to the white light curve is shown in Fig.~\ref{fig:raw_wlc}.

From the white-light curve fit, we measure $T_0$ = 60244.520379$_{-0.000014}^{+0.000015}$\,BJD, a planet-to-star radius ratio of $R_\mathrm{p}/R_\mathrm{s}=0.12199 \pm 0.00011$, a semi-major axis of $a/R_\mathrm{s}$ = 3.7868$_{-0.0098}^{+0.0097}$, and an impact parameter of $b$ = 0.150$_{-0.021}^{+0.0019}$. Our values for $a/R_\mathrm{s}$ and $b$ are consistent within 1-$\sigma$ of the constraints presented in \citet{delrez_wasp-121_2016} ($a/R_{s}$ = 3.754$_{-0.028}^{+0.023}$, $b$ = 0.160$_{-0.042}^{+0.040}$) and \citet{patel_empirical_2022} ($a/R_{s}$ = 3.81$_{-0.02}^{+0.01}$, $b$ = 0.091$_{-0.057}^{+0.052}$), but in slight disagreement with the measurements of \citet{bourrier_optical_2020} ($a/R_{s}$ = 3.8131$_{-0.0060}^{+0.0075}$, $b$ = 0.10$\pm$0.01). To ensure that our constraints on the orbital parameters are not driven by the presence of any eclipse mapping signal~\citep{challener_latitudinal_2024}, we also perform a fit where we discard all integrations more than 5 hours away from the mid-eclipse time and measure orbital parameters consistent at 1-$\sigma$ with those from the full phase-curve fit. 

We also perform the white-light curve fit allowing for eccentric orbits by fitting for $e\cos\omega$ ($\mathcal{U}[-1,1]$) and $e\sin\omega$ ($\mathcal{U}[-1,1]$). We find that the 3-sigma confidence interval of $e\cos\omega$ is constrained to $e\cos\omega$ = $-$0.00029--0 and measure a value of $e\sin\omega$ = $-$0.0085$_{-0.0013}^{+0.0012}$, corresponding to an eccentricity of $e$ = 0.0085$_{-0.0012}^{+0.0013}$ and an argument of periapsis that is constrained within 3-$\sigma$ to $\omega$ = 87.9--90.0$^\circ$. Because $e\cos\omega$ dictates the delay in eclipse time whereas $e\sin\omega$ dictates the difference in the transit and eclipse durations~\citep{winn_transits_2014}, our non-zero eccentricity measurement is thus driven by a difference in the duration of the eclipse relative to the primary transit. One possible explanation for this difference is our consideration of a uniform dayside when modelling the shape of the secondary eclipse signal. If the dayside distribution is non-uniform, for example due to morning/evening asymmetries, the fit could correct for this lack of flexibility in our secondary eclipse model by instead adjusting the eccentricity. Given this, as well as the fact that secondary eclipse spectra are not very sensitive to orbital parameters (contrary to transmission spectra), we opt to fix $e$ = 0 and $\omega$ = 90$^\circ$ for the spectroscopic light curves. When accounting for the Doppler boosting and ellipsoidal variation effects in the light curve, we measure 3-$\sigma$ upper limits on the amplitudes of $D<$116\,ppm and $E<$64\,ppm, respectively. Given that both of these measurements are consistent with 0, and that the expected values are negligible compared to the amplitude of the planetary signal~\citep[$D$ = 4\,ppm and $E$ = 17\,ppm over the TESS bandpass,][]{bourrier_optical_2020}, we set these to 0 for the spectroscopic light curve fits. 

Upon inspection of our phase curve model fit to the white-light curve, we find a preference for scenarios where the planetary flux dips to negative values near the primary transit, reaching a minimum of $-$103$\pm$9\,ppm at phase 170$^\circ$. We find that this negative flux remains when considering second- and third-order polynomials for the systematics model, with minimum flux values reaching down to $-$159$\pm$16. However, it is unlikely that this dip to negative planetary flux values affects the measured secondary eclipse spectrum given its phase separation to the minimum of the phase curve. We verify this by comparing our results with both the eclipse spectrum extracted with a penalization for unphysical negative fluxes from \cite{splinter_precise_2025}, and with spectra extracted for individual eclipses only considering baseline surrounding the eclipse (see Section~\ref{subsec:tests}).

\begin{figure}
    \centering
    \includegraphics[width=0.49\textwidth]{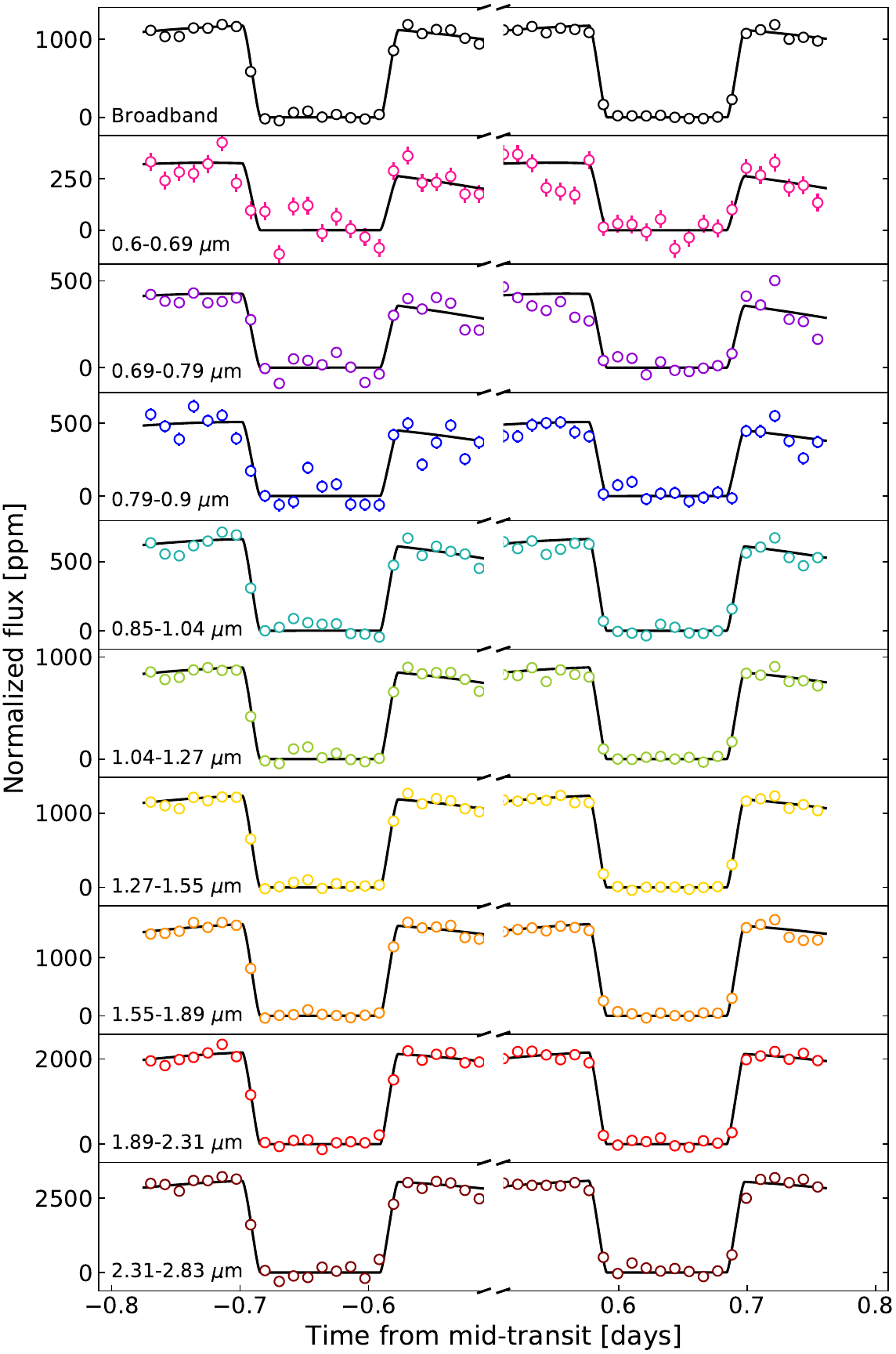}
    \vspace{-4mm}
    \caption{Broadband and spectroscopic secondary eclipses observed as part of the NIRISS/SOSS phase curve of WASP-121b, extracted with the \texttt{NAMELESS} pipeline. The light curve fit (black line) of the broadband order 1 light-curve ($\lambda=0.85-2.83\,\mu$m) is shown, along with the data binned at a resolution of 25 integrations per bin for visual clarity. The 122 light curves of the second order and 361 light curves of the first order are binned into 3 and 6 light curves, respectively (coloured points).
    }
    \label{fig:lightcurves}
    \vspace{-3mm}
\end{figure}

\subsubsection{Spectroscopic Light Curve Fit}\label{subsec:specLCF}

\begin{figure}
    \centering
    \includegraphics[width=0.49\textwidth]{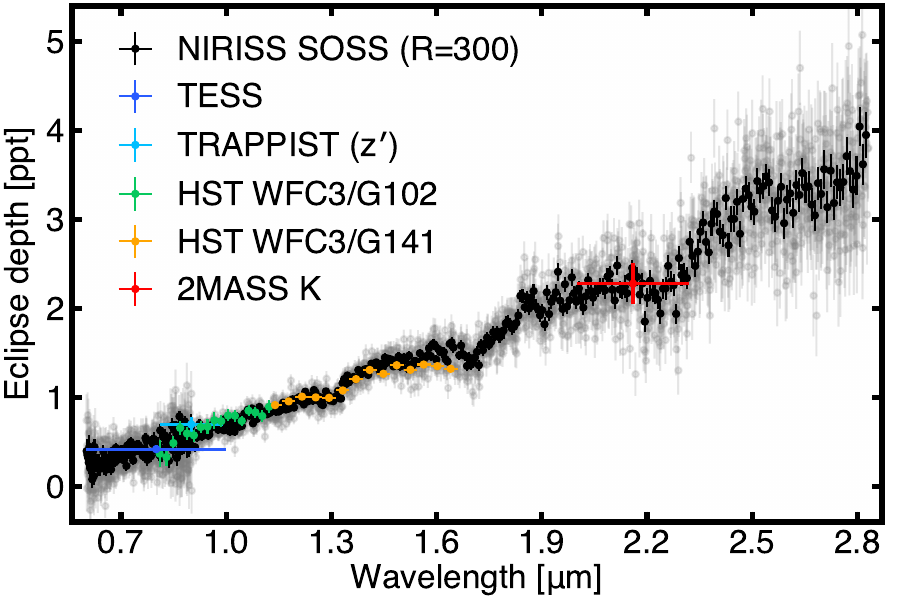}
    \vspace{-5mm}
    \caption{Extracted dayside planet-to-star flux ratio spectrum of WASP-121b observed with NIRISS SOSS, shown at both native resolution (grey points) and binned at a fixed resolving power of $R$ = 300 (black points). The NIRISS data are in relatively good agreement with previous observations obtained with TESS~\citep[][blue point]{bourrier_optical_2020}, TRAPPIST~\citep[][light blue point]{delrez_wasp-121_2016}, HST WFC3/G102~\citep[][green points]{mikal-evans_emission_2019}, HST WFC3/G141~\citep[][orange points]{mansfield_unique_2021}, and 2MASS~\citep[][red point]{kovacs_secondary_2019}, while also spanning a wider wavelength coverage.}
    \label{fig:eclipse_spec}
    \vspace{-3mm}
\end{figure}

We perform two spectroscopic light curve fits: one at a fixed resolving power of $R = 300$ and one at the pixel resolution (one light curve per detector column). To fit the spectroscopic light curves, we follow the same methodology as before, but now fixing the orbital parameters to the best-fit values of the white-light curve fit. The systematics model is applied in the same manner as for the white-light curve fit, allowing for the slopes and tilt-event jump amplitudes to vary with wavelength. The white-light and spectroscopic light curve fits are shown in Fig.~\ref{fig:lightcurves}. From these fits, we then extract the planet-to-star flux ratio values at mid-eclipse for all wavelength bins, producing the eclipse spectrum of WASP-121b (Fig.~\ref{fig:eclipse_spec}). The median and $\pm1\sigma$ uncertainties of the eclipse spectrum are computed via the samples of $F_n$ and $\delta_n$ in Eq.~\ref{eq:eclipse_depth_data} to determine the corresponding range of $F_p/\bar{F_s}$ values at $t=T_\mathrm{sec}$.

\subsection{Consistency tests}\label{subsec:tests}

To ensure the robustness of any inferred results on the reduction and fitting methods, we also use the independent extraction of the WASP-121b NIRISS eclipse spectrum done with the \texttt{exoTEDRF} pipeline~\citep{feinstein_early_2023, radica_awesome_2023, radica_muted_2024, radica_exotedrf_2024} from \cite{splinter_precise_2025}, to which we refer the reader for a full description of the data extraction and light curve fitting. Rather than at a constant resolving power, this reduction uses 5-pixel bins. Although the ensuing retrieval analysis focuses on the \texttt{NAMELESS} reduction, both reductions produce consistent spectra and atmospheric constraints (see Appendix~\ref{sec:jared_reduction}).

We also tested fitting both observed eclipses separately, treating each as if it had been observed as a single-eclipse configuration (with only a few hours of baseline before ingress and after egress). 
The single-eclipse extracted spectra show some differences in average depth at redder wavelengths, but otherwise produce constraints on the composition and thermal profile parameters that are mostly consistent with each other, and with the joint eclipse fit. The analysis of these is detailed in Appendix~\ref{sec:individual_eclipses}.

\section{Atmospheric Modelling} \label{sec:modelling}

To generate synthetic spectra of WASP-121b we use the SCARLET atmosphere modelling framework~\citep{benneke_atmospheric_2012, benneke_how_2013,  benneke_strict_2015, benneke_sub-neptune_2019, benneke_water_2019, pelletier_where_2021}. Our models consider molecular cross sections computed using \texttt{HELIOS-K}~\citep{grimm_helios-k_2015, grimm_helios-k_2021} for H$_2$O~\citep{polyansky_exomol_2018}, CO~\citep{rothman_hitemp_2010, li_rovibrational_2015}, OH~\citep{rothman_hitemp_2010}, SiO~\citep{yurchenko_exomol_2022}, CH$_4$~\citep{hargreaves_accurate_2020}, CO$_2$~\citep{yurchenko_exomol_2020}, HCN~\citep{barber_exomol_2014,harris_improved_2006}, VO~\citep{mckemmish_exomol_2016}, TiO~\citep{mckemmish_exomol_2019}, and FeH~\citep{wende_crires_2010}. For the atomic species Na and K, we use the opacities from the \texttt{petitRADTRANS}~\citep{molliere_petitradtrans_2019} database, which include the latest wing profiles~\citep{allard_kh2_2016, allard_new_2019}. For bound-free (H$^{-}_{\mathrm{bf}}$) and free-free (H$^{-}_{\mathrm{ff}}$) components of the H$^{-}$ opacity that is important in ultra-hot Jupiter atmospheres~\citep[e.g,.][]{vaulato_hydride_2025}, we compute these analytically following \cite{gray_observation_2021}. Here H$^{-}_{\mathrm{bf}}$ directly depends on the abundance of H$^{-}$ anions while H$^{-}_\mathrm{ff}$ depends on the abundance of both neutral hydrogen atoms and free electrons. Generated models also consider collision-induced absorption from H$_2$–H$_2$ and H$_2$–He interactions~\citep{borysow_collision-induced_2002}.

Given a vertical temperature structure and abundance profiles for each chemical species, SCARLET computes the top-of-atmosphere thermal flux emitted by WASP-121b ($F_p$).  We model the atmosphere as a grid of 50 pressure levels uniformly spaced in log between 10\,bar and 10$^{-6}$\,bar.  For our main analysis, models are computed at a spectral resolution of $R$ = 15,625, although we also tested higher resolving powers of $R$ = 31,250 and $R$ = 62,500, finding that these produce nearly identical results. The choice to not generate models at a higher spectral resolution than needed is mainly to avoid unnecessary use of computational resources when sampling many millions of models during atmospheric retrievals. Generated models of WASP-121b are then broadened with a rotational kernel assuming tidally locked synchronous rotation with a limb velocity of 6.99\,km\,s$^{-1}$.

To model the star, we use a PHOENIX template~\citep{husser_new_2013} interpolated from a grid to match the parameters of WASP-121~\citep[T$_\mathrm{eff}$ = 6459\,K, $\log$g = 4.242, and {[Fe$/$H]} = 0.13,][]{delrez_wasp-121_2016}. We Doppler shift the model at the relative radial velocity between target and instrument at the time of the observations considering both the systemic velocity of WASP-121~\citep{borsa_atmospheric_2021} and the motion of JWST relative to the barycentre. We convolve the stellar spectrum ($F_s$) with a rotational broadening kernel assuming a vsin$i$ = 13.5\,km\,s$^{-1}$~\citep{delrez_wasp-121_2016} and use this to compute the planet-to-star contrast ratio~\citep[$R_\mathrm{star}$ = 1.458\,$R_\odot$, $R_p$ = 1.753\,$R_J$,][]{bourrier_optical_2020}. The choice of using a template as opposed to the observed star-only spectrum during eclipse is to be able to model the eclipse spectrum at high resolution and to avoid introducing systematic errors from imprecisions in absolute flux calibration of the NIRISS data~\citep{coulombe_broadband_2023}.

With the NIRISS wavelength coverage extending down to the optical, reflected light can potentially significantly affect the measured eclipse depth at shorter wavelengths~\citep[e.g.,][]{schwartz_balancing_2015, keating_revisiting_2017, coulombe_highly_2025}, depending on the light back scattering efficiency of the dayside atmosphere of WASP-121b. To account for this, we also include the option of adding a reflected light contribution to the modelled eclipse depth spectrum,
\begin{equation} \label{eq:eclipse_depth_model}
    \mathrm{eclipse~depth} = A_{\mathrm{HS}}\frac{F_p}{F_s} \left( \frac{R_p}{R_s} \right)^2 + A_g  \left( \frac{R_p}{a} \right)^2,
\end{equation}
where $A_{\mathrm{HS}}$ is a hot spot area multiplicative factor~\citep{coulombe_broadband_2023}, $A_g$ is the geometric albedo, here assumed to be constant over the NIRISS bandpass, and $a$ is the planet-star distance. As emitted thermal flux scales with temperature to the fourth power,  $A_{\mathrm{HS}}$ is included in order to account for the possibility that not all regions of the observable planet contribute equally to the total emission spectrum~\citep{taylor_understanding_2020}. For example, the combination of a hotspot near the substellar point could result in an effective emitting area that is smaller than the size of WASP-121b as measured from transit observations. For WASP-121b, $(R_p/a)^2 \sim 995$\,ppm, meaning that even a relatively low but non-zero albedo can affect the eclipse depth by tens to hundreds of parts per million which, at the precision of the NIRISS WASP-121b spectrum, can have a large impact for fitting the bluest wavelength data points.

\subsection{Retrieval Setup} \label{subsec:retrieval}

In this work we explore a range of atmospheric retrieval prescriptions to characterise the dayside atmosphere of WASP-121b.  To fit for the temperature structure, we use the flexible parameterization of \cite{pelletier_where_2021} based on the approach of \cite{line_uniform_2015}.  In our case this consists of fitting 12 temperature `knots' uniformly distributed in log pressure between 10 and 10$^{-6}$\,bar with a smoothing prior of $\sigma_{\mathrm{smooth}}$ = 500\,K\,dex$^{-2}$ on the second derivative to prevent non-physical temperature oscillations in pressure regions not well constrained by the data.

While the opacity floor of the deepest layers of the dayside atmosphere of ultra-hot Jupiters is expected to be set by H$^{-}$, we also allow our models the additional flexibility of fitting for an achromatic continuum level at a given $\log$ pressure level ($\log_{10} P_c$, $\log$bar).  Practically this can mimic either the presence of a grey continuum absorber such as a cloud deck, or the temperature-pressure (TP) profile becoming isothermal below this level. As in \cite{coulombe_broadband_2023} we also fit for an area fraction factor ($A_{\mathrm{HS}}$) multiplied to the thermal emission spectrum (Eq.~\ref{eq:eclipse_depth_model}). In some retrievals we also freely fit for the geometric albedo ($A_g$). In the case that error bars may be underestimated, we also fit for a multiplicative error inflation term to the NIRISS data points. For all our retrievals we use \texttt{emcee}~\citep{foreman-mackey_emcee_2013} as a sampler.

For the characterisation of exoplanetary atmospheres, commonly used retrieval prescriptions can range from making no underlying chemical assumptions, to modelling the atmosphere in full chemical equilibrium, with either methods and anything in between having its advantages and caveats. As ultra-hot Jupiter atmospheres are particularly extreme environments, it is not necessarily obvious a priori what retrieval prescription is optimal to accurately infer their atmospheric properties. On the one hand, retrievals that assume the atmosphere to be well mixed will miss key physics such as thermal dissociation which are expected to play a significant role in shaping abundance profiles of certain species in these conditions.  On the other hand, retrievals assuming chemical equilibrium may miss disequilibrium processes such as cold-trapping that can also lead to biases in inferred atmospheric characteristics. We now detail the different retrieval types explored in this work. The motivation for utilising different retrieval prescriptions is to obtain a consistent picture of the atmospheric properties of WASP-121b that is independent of the modelling assumptions made.

\subsection{Free retrievals}\label{subsubsec:free_retrieval}

The first class of atmospheric retrievals that we explore are `free' or `well mixed' retrievals which consist of parameterizing the $\log_{10}$ volume mixing ratio (VMR) of all included opacity sources as being constant with altitude, each as a free parameter fitted in the retrieval. For fitting the NIRISS dayside spectrum of WASP-121b, we include H$_2$O, CO, OH, CH$_4$, CO$_2$, HCN, VO, TiO, SiO, FeH, Na, K, H$^{-}$, and e$^{-}$. The purpose of such an approach employing many free abundance parameters is to have a more data driven approach to what best fits the observed spectrum, regardless of whether it makes sense physically (e.g., an atmosphere made of 50\% CH$_4$ and 50\% VO is allowed, even though this would not make sense considering the density and temperature of WASP-121b, or the low natural abundance of V atoms in the Universe), and then interpreting this in the context of chemistry and measured elemental abundances from the Sun or the host star.

With queried abundances for each included chemical opacity contributor, the retrieval then sums the abundances of all included species and then adds H$_2$, He, and H as `filler gases' such that the volume mixing ratio of all species included always add to one in every atmospheric layer. The relative amounts of H$_2$, He, and H added depend on the temperature-pressure profile and is determined using \texttt{FastChemCond}~\citep{stock_fastchem_2018, stock_fastchem_2022, kitzmann_fastchem_2024}, with H$_2$ typically being most abundant deeper in the atmosphere and H dominating at higher altitudes for ultra-hot Jupiter atmospheres \citep[e.g.,][see their Extended Fig.~6]{pelletier_vanadium_2023}. The calculated H abundance profile is also later multiplied to the electron pressure to calculate the H$_{\mathrm{ff}}$ opacity. In this sense the retrieval does utilise some knowledge of equilibrium chemistry, but only for the scale height and H$_{\mathrm{ff}}$ calculations. Accounting for the transition between atomic and molecular hydrogen is important as it corresponds to a change in atmospheric mean molecular weight (and hence scale height) by nearly a factor of two. For dayside of WASP-121b, the H$_2$ to H transition typically occurs around 10\,mbar in our models. This is particularly relevant for fitting data with multiple strong spectral bands that probe a wide range of atmospheric pressures.  For example, H$_2$O (which is preferentially dissociated at higher altitudes) probes deeper pressure levels that may still be in the H$_2$-dominated chemical regime. Contrastingly, species such as CO and VO that are more stable to dissociation and/or have stronger opacity bands can probe down to lower pressures in the puffier H-dominated regions of the atmosphere. This large change in mean molecular weight between the lower (H$_2$ dominated) and upper (H dominated) parts of the atmosphere, in turn, can alter the strength and shape of spectral lines, although this effect will be particularly important for accurately modelling transmission spectra~\citep{pluriel_strong_2020, gapp_wasp-121_2025}.  

As in \cite{coulombe_broadband_2023, pelletier_crires_2025}, we also test `hybrid' free retrievals that are functionally the same as classical free retrievals, but that have abundance profiles for certain species that can vary with altitude following \citet{parmentier_thermal_2018}.  Specifically, all species included in our retrievals have constant-with-altitude abundances in the deep atmosphere, except for H$_2$O, TiO, VO, Na, K, and H$^{-}$ which are parameterized as having decreasing abundance following a power law at a cutoff pressure~\citep[][see their Table 1]{parmentier_thermal_2018}. Accounting for thermal dissociation of molecules is particularly important for ultra-hot atmospheres and can otherwise lead to significant biases in inferred abundances~\citep{gandhi_revealing_2024, pelletier_crires_2025, bazinet_quantifying_2025}.

\begin{table}
\caption{Atmospheric retrieval parameter description and prior ranges for the free and chemical equilibrium retrievals.} 
\vspace{-3mm}
\label{tab:retrieval_priors}
\centering
\def\arraystretch{1.1}
\begin{tabular}{ccc}
\hline
\hline
Parameter & Description & Prior range \\
\hline
\hline
$T_i  $ & Temperature of layer $i$ [K] & $\mathcal{U}$[100,6000] \\
$A_g  $ & Geometric albedo & $\mathcal{U}$[0,1] \\
$A_{\mathrm{HS}}$ & Hotspot area factor & $\mathcal{U}$[0,1] \\
$\log P_\mathrm{c}$ & Continuum pressure [$\log\mathrm{bar}$] & $\mathcal{U}$[$-$5,1] \\
Error inf.\ & Error inflation factor & $\mathcal{U}$[1,5] \\
\hline
 & Free and hybrid free & \\
\hline
$\log$VMR$_j$ & VMR of species $j$ & $\mathcal{U}$[$-$12,0] \\
\hline
 & Chemical equilibrium (single) & \\
\hline
$[$M$/$H$]$ & Global $\log$ metallicity & $\mathcal{U}$[$-$3,3] \\
C$/$O & Carbon-to-oxygen ratio & $\mathcal{U}$[0,10] \\
\hline
 & Chemical equilibrium (multi) & \\
\hline
$[\mathcal{V}/$H$]$ & Volatile $\log$ metallicity & $\mathcal{U}$[$-$3,3] \\
$[\mathcal{R}/$H$]$ & Refractory $\log$ metallicity & $\mathcal{U}$[$-$3,3] \\
$[$Ti$/$H$]$ & Titanium $\log$ metallicity & $\mathcal{U}$[$-$3,3] \\
C$/$O & Carbon-to-oxygen ratio & $\mathcal{U}$[0,10] \\
\hline
\multicolumn{3}{l}{\small $i$ = 0 to 11 ($T_{0}$ at $10^{-6}$\,bar and $T_{11}$ at $10$\,bar)}\\
\multicolumn{3}{l}{\small $j$ =  H$_2$O, CO, OH, CH$_4$, CO$_2$, HCN, VO, TiO SiO, FeH, Na, K,}\\
\multicolumn{3}{l}{\small H$^{-}$, and e$^{-}$}
\vspace{-3mm}
\end{tabular}\\ 
\end{table}

\subsection{Chemical equilibrium retrievals}\label{subsubsec:chem_retrieval}

Another flavour of atmospheric retrievals that we explore are `chemical equilibrium' retrievals. Rather than fitting every absorber as a free parameter, these typically assume a C$/$O ratio and a global metallicity, with other elemental abundances held in relative solar proportions and abundance profiles for all included species instead calculated from a Gibbs free energy minimisation chemical network, in our case using \texttt{FastChemCond}. 

Chemical equilibrium retrievals have the benefit of naturally incorporating physical processes such as thermal dissociation and ionisation of species which can give rise to highly non-uniform abundance profiles.  However, they generally lack the flexibility of exploring compositions outside those allowed by equilibrium chemistry (e.g., due to vertical mixing or cold-trapping), or that have different relative elemental abundances than those assumed in \texttt{FastChemCond}. For example, if a species is missing from the gas phase due to a cold-trap~\citep[e.g.,][]{pelletier_vanadium_2023}, a chemical equilibrium network would overpredict its abundance and likely bias any retrieved atmospheric properties. Additionally, assuming that all other metals except C relative to O (which the retrieval can vary via the C$/$O ratio) are held in solar proportions may be an erroneous assumption given that giant planets do not necessarily maintain solar-like volatile-to-refractory ratios during formation~\citep{lothringer_new_2021, turrini_tracing_2021, chachan_breaking_2023}.

\begin{figure*}[ht!]
    \centering
    \includegraphics[width=\textwidth]{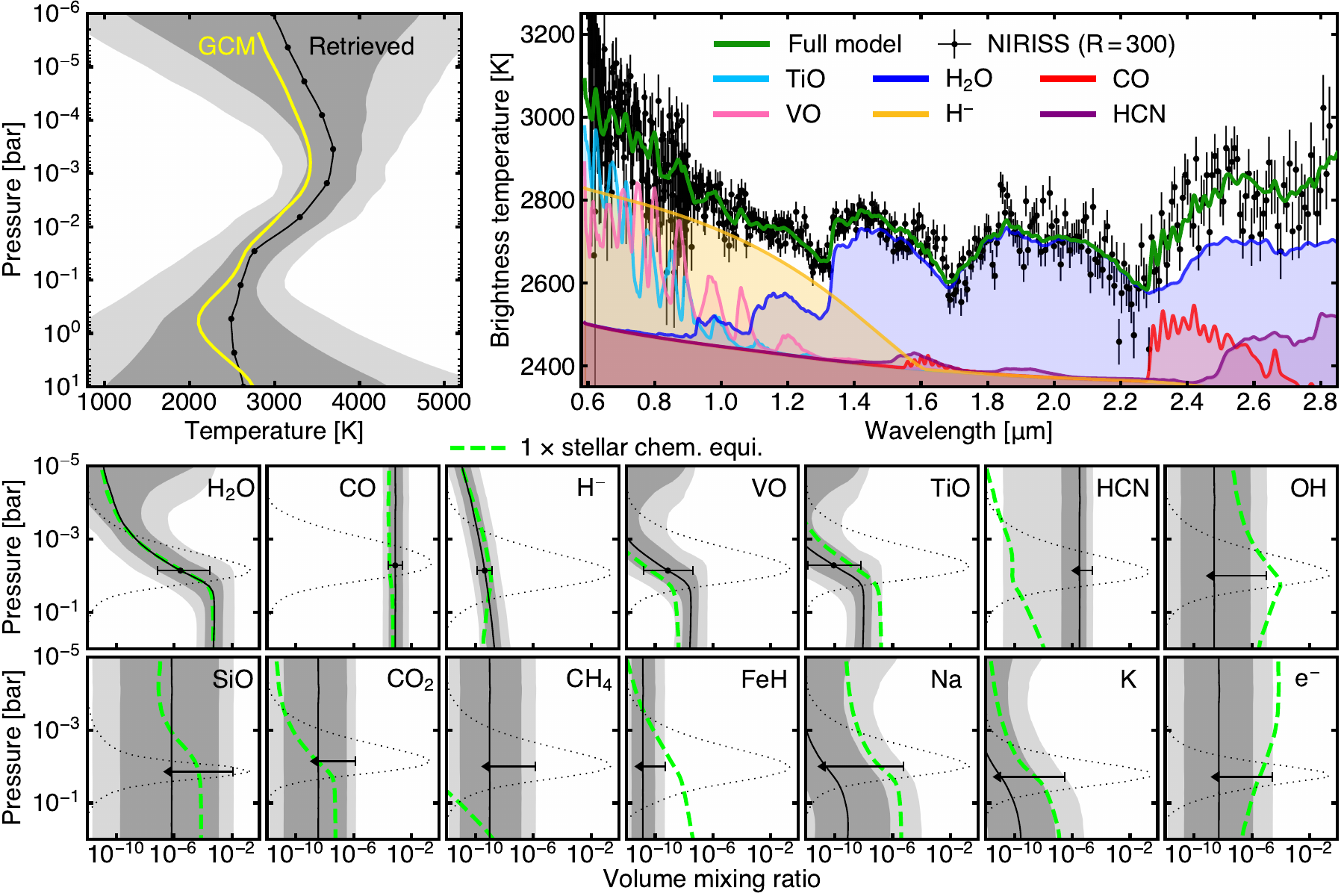}
    \vspace{-5mm}
    \caption{
    Results overview for the WASP-121b NIRISS SOSS eclipse hybrid free retrieval including parameterized thermal dissociation.
    Top left: Retrieved dayside vertical temperature structure (black line: median, grey contours: 1 and 2 $\sigma$ bounds) compared to the average TP profile at mid-eclipse of a fiducial GCM model~\citep[][yellow line]{parmentier_thermal_2018}. 
    Top right: Measured dayside brightness temperature of WASP-121b (black points) compared to the best fit atmospheric model (solid green line).  Opacity contributions from individual species are shown in colour.  Multiple H$_2$O emission bands can be seen, as well as contribution from CO around 2.4\,$\mu$m.  At shorter wavelengths, H$^{-}$, VO, and TiO are necessary to match the rise in brightness temperature (note that an albedo of zero was assumed for this retrieval).
    Bottom: Each panel shows the retrieved abundance profiles for an individual species fitted in the retrieval (solid black line: median, grey contours: 1 and 2 $\sigma$ bounds).  Average contribution functions (dotted black line) show the pressure level probed by each species. The abundance profiles predicted from an equilibrium chemistry model assuming the median TP profile and a stellar-like composition~\citep{evans-soma_sio_2025} are also shown for comparison (dashed lime green lines). Constrained abundances are obtained for H$_2$O, CO, VO, TiO, and H$^{-}$, with only upper limits being obtained for the other species. Error bars (1$\sigma$) or upper limits (2$\sigma$) are shown at the approximate average pressure probed (peak of the contribution function).
    }
    \label{fig:free_retrieval}
    \vspace{-3mm}
\end{figure*}

To alleviate potential issues arising from fitting for a single atmospheric metallicity, we also explore chemical equilibrium retrievals with distinct abundance parameters controlling different group of chemical species, similarly as in \cite{pelletier_crires_2025}. This has the advantage of still including equilibrium chemistry effects such as thermal dissociation while also allowing for atmospheric scenarios with non-solar abundance ratios (e.g., due to cold-trapping). For this, we bundle volatile species together as a ``volatile metallicity'' [$\mathcal{V}/$H], and refratories species together as a ``refractory metallicity'' [$\mathcal{R}/$H], with the exception of titanium which we fit as its own parameter [Ti$/$H] due to its ultra-refractory nature~\citep{lodders_solar_2003}. Here the notation [$\mathcal{V}/$H] refers to the $\log_{10}$ metallicity relative to solar~\citep{asplund_chemical_2009}, with e.g., a value of [$\mathcal{V}/$H] = 1 meaning that volatiles are enriched relative to solar by a factor of ten. For a given queried C$/$O ratio, the retrieval thus has [$\mathcal{V}/$H] controlling the enrichment of volatile elements C, N, and O (setting the abundances of H$_2$O, CO, OH, CH$_4$, CO$_2$, HCN, H$^{-}$ and e$^{-}$), [$\mathcal{R}/$H] controlling the enrichment of refractory elements V, Si, Fe, Na, K (setting the abundances of VO, SiO, FeH, Na, and K), and [Ti$/$H] controlling enrichment of the ultra-refractory element Ti (setting the abundances of TiO). While only the listed species are considered as line absorbers in our model, abundance profiles computed with \texttt{FastChemCond} consider a full chemical network. For example, [Ti$/$H] considers Ti, Ti$^+$, TiO$_2$, TiH, etc when predicting the TiO abundance for a given TP profile.

\section{Results and Discussion}\label{sec:results}

From the extracted NIRISS brightness temperature spectrum of WASP-121b, multiple water bands can be clearly identified by eye, in particular those at 1.2, 1.5, and 2.0\,$\mu$m (Fig.~\ref{fig:free_retrieval}, top right panel). The water bands are notably contained between brightness temperatures of about 2600\,K and 2800\,K and in emission, indicating a rise of 200\,K with increasing altitude over the pressure range probed by H$_2$O. Additional emission is also apparent around 2.4\,$\mu$m where CO has a bandhead that was previously observed on WASP-121b using ground-based high-resolution spectroscopy~\citep{smith_roasting_2024, pelletier_crires_2025}. The higher brightness temperature of $\sim$2900\,K at these wavelengths is consistent with WASP-121b's thermal inversion extending to higher temperatures at higher altitudes and CO probing lower pressures than H$_2$O on average.  This qualitative picture also matches well the H$_2$O and CO features observed in the NIRSpec data~\citep[][see their Figure 1, bottom left panel]{evans-soma_sio_2025}. At bluer wavelengths, the steep rise in emission below $\sim$1\,$\mu$m indicates either the presence of strong optical opacity (e.g., due to H$^{-}/$TiO$/$VO) combined with a stratosphere extending to at least $\sim$3100\,K at lower pressures than probed by H$_2$O and CO, or a non-negligible reflected light contribution.

In the following subsections, we explore a range of retrieval prescriptions to study what chemical components and physical processes are needed to best fit the NIRISS data and infer the properties of WASP-121b's dayside atmosphere. Overall we find that H$_2$O, CO, VO, and H$^{-}$ have bounded abundance constraints in all retrievals tested. The inclusion of H$_2$O dissociation is not strictly required to fit the spectrum, but strongly biases the inferred abundances if neglected.  The bluest wavelengths can be fitted by either TiO or reflected light, but in all cases TiO is underabundant relative to other species, possibly due to partial cold-trapping. Given the underabundance of TiO, assuming chemical equilibrium and calculating abundances from a single overarching metallicity will overpredict the TiO abundance and bias all other inferred parameters. Meanwhile the volatile-to-refractory ratio is consistent with the solar and stellar values.

\subsection{Free retrieval atmospheric exploration}\label{free_retrieval_results}

As a first test, we run a hybrid free retrieval that includes parameterized dissociation abundance profiles, but which assumes an albedo of zero to explore what models best fit the spectrum assuming only a planetary thermal emission component. In tandem to the atmospheric composition, we simultaneously retrieve the TP structure of the dayside atmosphere of WASP-121b, finding it to have a strong thermal inversion similar to the mean hemispheric profile at eclipse phase from the fiducial solar-composition GCM of \cite{parmentier_thermal_2018} (Fig.~\ref{fig:free_retrieval}, top left panel). We find that, under the assumption of no reflected light, contributions from a combination of H$_2$O, CO, VO, TiO, and H$^{-}$ are needed to best match the overall shape of the spectrum (Fig.~\ref{fig:free_retrieval}, top right panel).  

As free retrievals can allow for unphysical atmospheric abundances, we compare our retrieved abundance profiles (Fig.~\ref{fig:free_retrieval}, bottom panels) to equilibrium chemistry predictions computed assuming the median retrieved TP profile and a stellar-like composition~\citep{evans-soma_sio_2025}. While the composition of WASP-121b will not necessarily equal that of its host star, this can act as a good first comparison of how the inferred composition differs from this reference point. The retrieved H$_2$O, CO, and H$^{-}$ abundance posteriors are consistent with the fiducial chemical equilibrium predictions (dashed lime green lines in Fig.~\ref{fig:free_retrieval}). Interestingly, while VO is slightly above the reference stellar composition equilibrium model, TiO appears marginally less abundant than the equilibrium chemistry prediction. This is somewhat analogous to the atomic Ti signal being weaker than expected compared to the atomic V signal seen in WASP-121b's transmission spectrum at high resolution~\citep{prinoth_titanium_2025}. We further note that while VO bands are visible in the spectrum around 0.95 and 1.05\,$\mu$m, TiO serves more to fit the steep rise in opacity at the bluest wavelengths rather than to match any noticeable TiO bands. Indeed the posterior on the TiO abundance is no longer bounded if we allow for a reflected light contribution in the retrieval (see Section~\ref{subsubsec:titanium}), hence we do not consider this to be a detection of TiO.

In some of the posterior space, HCN partly contributes to fitting the red end of the spectrum, with a maximum in the probability distribution at VMR$_{\mathrm{HCN}} = 10^{-6}$. However, with a posterior tail extending to the lower prior bound, this only constitutes a weak preference for HCN rather than a detection. Furthermore, while other posteriors remain consistent, no evidence of any HCN contribution is found if performing the same retrieval on the \texttt{exoTEDRF} reduction, likely indicating that this is an artifact driven by the few elevated data points at the noisy red end of the NIRISS detector in the \texttt{NAMELESS} reduction. While OH and SiO have previously been reported on the dayside atmosphere of WASP-121b~\citep{smith_roasting_2024, evans-soma_sio_2025}, the NIRISS data has limited sensitivity to these species, with only broad upper bounds being inferred that cannot confirm or rule out their presence. For OH, this is due to the lack of broad molecular bands in the OH spectrum and the relatively low spectral resolving power of NIRISS that cannot resolve individual spectral lines. For SiO, this is rather because of the lack of any strong spectral features in the NIRISS bandpass. Only upper limits that encompass the expected abundances predicted by equilibrium chemistry for a star-like composition are obtained for CO$_2$, CH$_4$, FeH, Na, K, and e$^{-}$.

\begin{figure}
    \centering
    \includegraphics[width=0.49\textwidth]{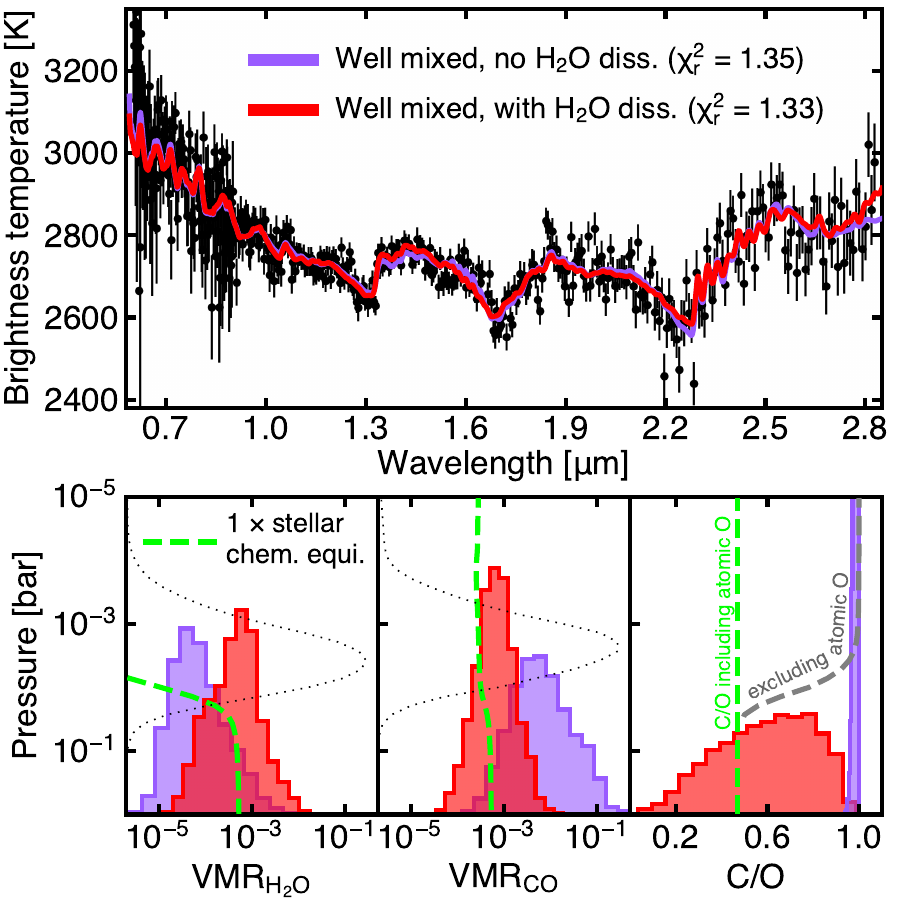}
    \vspace{-5mm}
    \caption{
    Importance of accounting for thermal dissociation in atmospheric composition inferences of ultra-hot Jupiters.   
    Top: Brightness temperature NIRISS spectrum (black points) overplotted with the best fit model from free retrieval analyses without (purple) and with (red) thermal dissociation parameterized. 
    Bottom: Inferred atmospheric compositions for H$_2$O (left), CO (middle), and the derived C$/$O ratio from counting atoms in all C- and O-bearing molecules included in the retrieval (right) compared to abundance profiles for a fiducial WASP-121b chemical equilibrium model assuming a stellar-like composition (dashed lime green lines). While CO is relatively constant with altitude, H$_2$O molecules are expected to be significantly thermally dissociated at pressures probed (dotted black lines). While the true atmospheric C/O ratio is constant at all pressures (dashed lime green line), the C/O derived from molecules only (dashed grey line) is not owing to the majority of oxygen atoms being in atomic form at lower pressures. As a result, the observed C$/$O ratio from only considering molecules will be biased to higher values (near unity) due to not accounting for atomic oxygen, which these observations are not sensitive to. Accounting for dissociation does lead to less precise but likely more accurate abundance measurements as it fits for the H$_2$O abundance of the deep atmosphere. Although the spectrum can still be well fitted when neglecting H$_2$O dissociation, the retrieved abundances and C$/$O ratio may not reflect the true bulk atmospheric composition.    
    }
    \label{fig:water_diss}
    \vspace{-3mm}
\end{figure}

\subsection{The impact of neglecting water dissociation}\label{h2o_diss}

\begin{figure*}
    \centering
    \includegraphics[width=\textwidth]{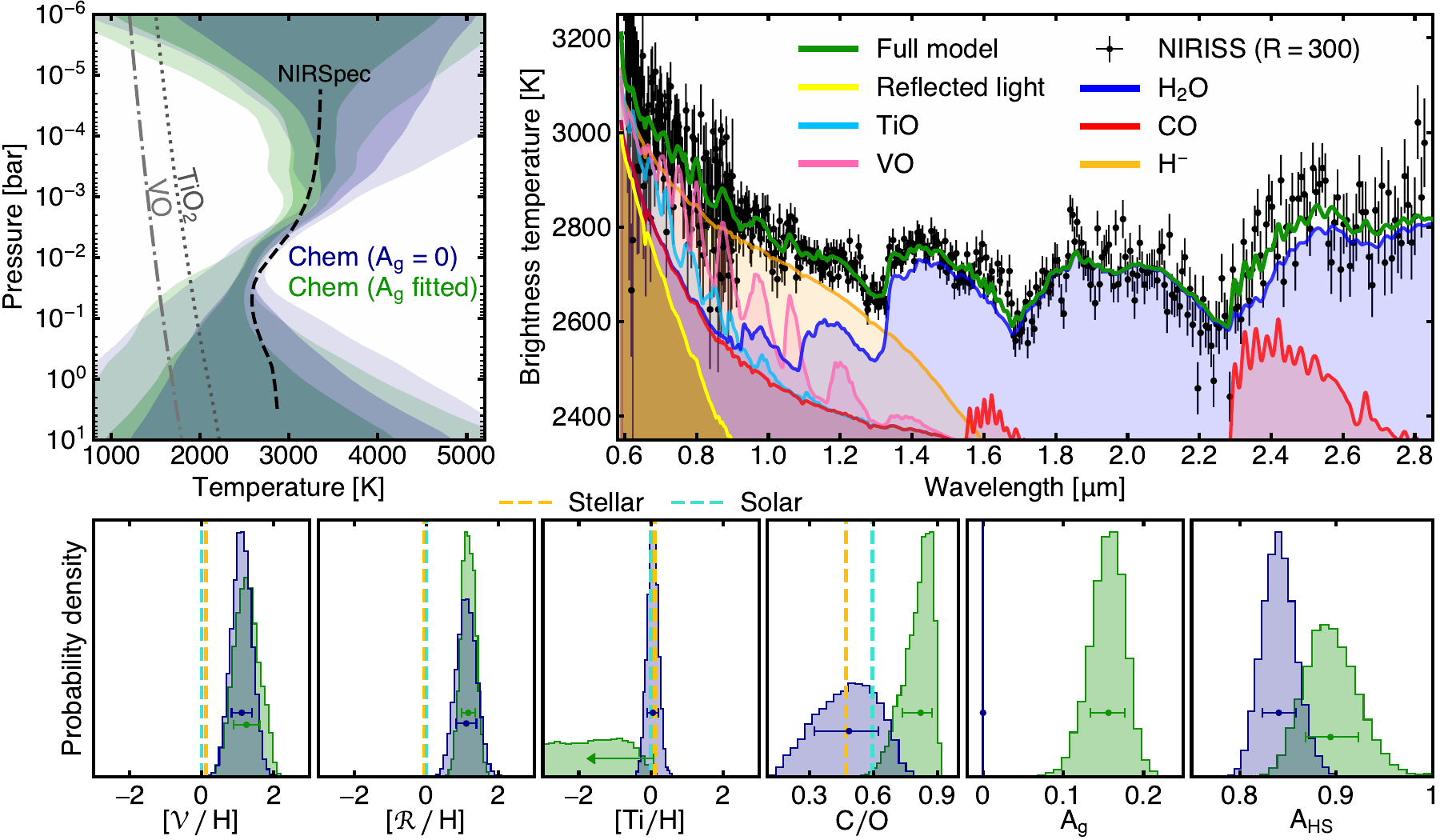}
    \vspace{-5mm}
    \caption{
    WASP-121b NIRISS SOSS eclipse chemical equilibrium retrieval results overview with (green) and without (dark blue) fitting for an albedo.  
    Top left: Retrieved dayside vertical temperature structure (1 and 2\,$\sigma$ contours) compared to the median TP profile inferred from NIRSpec G395H observations~\citep[][dashed black line]{evans-soma_sio_2025}. 
    While the temperature is generally above the VO (dashed dotted light grey line) and TiO$_2$ (dotted dark grey line) condensation curves~\citep{burrows_chemical_1999} on the dayside, the colder nightside may allow for some titanium condensation.
    Top right: Measured dayside brightness temperature of WASP-121b (black points) compared to the best fit atmospheric model from the chemical equilibrium retrieval with $A_g$ fitted (solid green line).  Opacity contribution from individual species and reflected light are shown in colour. Compared to when assuming $A_g = 0$ (Fig.~\ref{fig:free_retrieval}, top right panel), TiO is no longer necessary to match the data at shorter wavelengths when reflected light is considered.
    Bottom: From left to right, the log volatile metallicity, log refractory metallicity, log titanium metallicity, carbon-to-oxygen ratio, geometric albedo, and hotspot area scaling factor. Shown error bars are either 1\,$\sigma$ bounds or 3\,$\sigma$ upper limits.  While [$\mathcal{V}/$H] and [$\mathcal{R}/$H] are always above the stellar (orange dashed line) and solar (turquoise dashed line) values, the inference of [Ti$/$H], the C$/$O ratio, and $A_{\mathrm{HS}}$ depends on whether or not $A_g$ is fitted. In particular, the model either needs reflected light and an albedo of $A_g = 0.16_{-0.02}^{+0.02}$, or the combination of TiO and a slightly stronger temperature inversion to fit the spectrum. However, in both cases [Ti$/$H] is underabundant compared to other, less refractory metals.
    }
    \label{fig:chem_retrieval}
    \vspace{-3mm}
\end{figure*}

Here we compare the differences between fitting the observed NIRISS eclipse spectrum of WASP-121b using a free retrieval with constant-with-altitude abundance profiles (not accounting for dissociation) and a hybrid free retrieval that includes a parameterization for thermally dissociated abundance profiles (Section~\ref{subsubsec:free_retrieval}). Given that H$_2$O dissociation is expected to be important in the dayside atmosphere of WASP-121b~\citep{parmentier_thermal_2018, mikal-evans_emission_2019, smith_roasting_2024, pelletier_crires_2025, evans-soma_sio_2025, bazinet_quantifying_2025}, one might naively expect the latter to provide a significantly better fit to the data. However, we find that the NIRISS data can be fitted similarly well (reduced $\chi^2$ of 1.35 free versus 1.33 hybrid free) with both approaches (Fig.~\ref{fig:water_diss}, top panel). Calculating the simplified Bayesian Predictive Information Criterion (BPICS)~\citep{ando_bayesian_2007, ando_predictive_2011} following the recommendation of \cite{thorngren_bayesian_2025}, we find that the hybrid free retrieval is only favoured at the $\sim$2$\sigma$ level. Nevertheless, although only marginally improving the goodness of fit, the inclusion of thermal dissociation has a significant impact on inferred abundances, especially of H$_2$O relative to CO (Fig.~\ref{fig:water_diss}, bottom left panels). For the well mixed retrieval without H$_2$O dissociation, the constraints on the derived atmospheric C$/$O ratio are extremely narrow, between 0.98 and 1.0 (3$\sigma$ confidence). The strict bounds are set by the models preferring an atmospheric composition that is CO dominated (equal number of C and O atoms, C$/$O ratio = 1), with at best a much lower abundance of H$_2$O and other C- and O-bearing species relative to CO.  Rather than the atmosphere of WASP-121b being intrinsically water poor, this is most likely a result of these dayside observations probing lower pressure regions of the atmosphere where H$_2$O molecules are significantly dissociated. In the free retrieval, since all abundances are assumed to be constant with altitude (a decent approximation for CO but a poor one for H$_2$O in the conditions of WASP-121b's hot dayside), the retrieval compensates by adjusting the abundances of H$_2$O and CO (as well as the temperature structure and continuum level) to maintain the observed relative band strengths.

Effectively, what is happening is that the retrieval is measuring the abundances in the upper atmosphere and the C$/$O ratio is then inferred under the assumption that most of the C and O atoms are held in CO and H$_2$O (Fig.~\ref{fig:water_diss}, bottom right panel). Here H$_2$O dissociation produces a significant proportion of atomic O, which is not considered in the retrieval due to lack of spectral features. The C$/$O ratio calculated only from molecules without considering atomic oxygen will vary strongly with altitude and can lead to a biased estimation that is not representative of the deep atmosphere. Indeed while free retrievals can work well for warm and hot giant exoplanets that have CO and H$_2$O abundance profiles that do not vary significantly with altitude, they likely break down in the ultra-hot regime where thermal dissociation and ionisation become important~\citep{parmentier_thermal_2018, pluriel_toward_2022, brogi_roasting_2023, coulombe_broadband_2023, pelletier_crires_2025, evans-soma_sio_2025}.  

The importance of H$_2$O dissociation is made particularly evident by the detections of OH, a byproduct of H$_2$O thermal dissociation (alongside atomic O), in both the terminator and dayside regions of WASP-121b made using high-resolution spectroscopy~\citep{wardenier_phase-resolving_2024, smith_roasting_2024}. This means that any estimation of the planet's oxygen budget based on the measured H$_2$O abundance will likely not reflect that of the deep atmosphere. This can be especially problematic if attempting to link the retrieved atmospheric abundances to that of the bulk envelope, under the assumption that it reflects the primordial atmosphere which may hold imprints of the planet's formation and accretion history, for example via the C$/$O ratio. 
Including the effects of dissociation in a free retrieval does yield abundance constraints likely more representative of the deep atmosphere, in this case giving a relatively poorly constrained C$/$O ratio of 0.61$_{-0.25}^{+0.20}$ (Fig.~\ref{fig:water_diss}, bottom right panel).  While the large uncertainties may be more realistic, this also relies on the assumed parametric relation between the photosphere and higher pressures (Fig.~\ref{fig:free_retrieval}, middle left panel) which is likely not perfectly accurate.

\subsection{Chemical equilibrium approach}\label{chem_retrieval_results}

In this section we explore atmospheric retrievals that enforce equilibrium chemistry, with and without the inclusion of reflected light via a freely fit geometric albedo. In both cases, we retrieve a temperature structure that is similar to that found by \cite{evans-soma_sio_2025} using NIRSpec$/$G395H (Fig.~\ref{fig:chem_retrieval}, top left panel), although we note that the pressure onset of the thermal inversion is correlated to the metallicity. The thermal inversion extends to hotter temperatures at low pressures when the spectrum is fit assuming only planetary thermal emission ($A_g$ = 0). The inclusion of reflected light serves as an alternate means of fitting the excess emission at blue wavelengths (Fig.~\ref{fig:chem_retrieval}, top right panel), no longer requiring significant TiO opacity as in the case when the albedo was assumed to be null (Fig.~\ref{fig:free_retrieval}, top right panel). 

For the dayside composition of WASP-121b, we find the atmosphere to be metal rich in both volatile and refractory species whether or not reflected light is included (Table~\ref{tab:retrieval_results}). Contrastingly, [Ti$/$H] is measured to be underabundant relative to the other species in both cases, suggesting that it is partially depleted in even the optimistic $A_g = 0$ scenario (see Section~\ref{subsubsec:titanium}).  Compared to a stellar C$/$O value of $0.47 \pm 0.06$~\citep{evans-soma_sio_2025} which is slightly lower than the solar value of C$/$O = $0.59 \pm 0.05$~\citep{asplund_chemical_2021}, our inferred C$/$O ratio for WASP-121b ranges from being consistent with stellar in the case of zero albedo (C$/$O = $0.48_{-0.16}^{+0.14}$), or significantly super-stellar if reflected light is considered (C$/$O = $0.82_{-0.09}^{+0.05}$).  The dependence of the retrieved [Ti$/$H] and C$/$O ratio on the inclusion of reflected light is particularly striking (Fig.~\ref{fig:chem_retrieval}, bottom panels), and a cautionary tale for atmospheric inferences of ultra-hot Jupiters.  If $A_g = 0$, then the fit will adjust by increasing the strength of the inversion and adding extra optical opacity (TiO) to fit the observed rise in brightness temperature on the blue end of the spectrum. In turn, the stronger inversion has the effect of reducing the amount of CO needed to fit the 2.3\,$\mu$m bandhead, lowering the C$/$O ratio. The super-solar and super-stellar C$/$O = $0.82_{-0.09}^{+0.05}$ recovered when including an albedo is more in line with recent measurements of C$/$O = $0.70_{-0.10}^{+0.07}$~\citep{smith_roasting_2024} and C$/$O = $0.73_{-0.08}^{+0.07}$~\citep{pelletier_crires_2025} from ground-based high resolution spectroscopy, C$/$O between 0.59 and 0.87 from HST$/$WFC3~\citep{changeat_is_2024}, and C$/$O = $0.92_{-0.03}^{+0.02}$~\citep{evans-soma_sio_2025} from JWST$/$NIRSpec. In contrast to the free retrievals which can allow for C$/$O ratios up to unity, here the upper bound of C$/$O$\sim$0.92 (Fig.~\ref{fig:chem_retrieval}, C$/$O panel) comes from the chemistry prescription predicting a sharp change in composition reducing the abundance of O-only bearing molecules such as H$_2$O and VO in favour of producing more CO and C-only bearing molecules such as HCN and CH$_4$~\citep[e.g.,][]{madhusudhan_co_2012}. Only an upper limit was obtained for the grey continuum pressure (Table~\ref{tab:retrieval_results}).

When fitted, we retrieve an albedo $A_g = 0.16_{-0.02}^{+0.02}$, which is higher than expected considering both model predictions~\citep[e.g.,][]{malsky_direct_2024} and previous geometric albedo estimates of other ultra-hot Jupiters, which tend to be quite dark (non-reflective) at optical wavelengths~\citep{bell_very_2017, shporer_tess_2019, blazek_constraints_2022, demangeon_asymmetry_2024, deline_dark_2025}.  For WASP-121b specifically, previous estimates of its geometric albedo range from $0.16 \pm 0.11$~\citep{mallonn_low_2019} in the $z'$ band, to $0.07_{-0.040}^{+0.037}$~\citep{daylan_tess_2021} and $0.26 \pm 0.06$~\citep{wong_systematic_2020} from TESS. We note, however, that albedo estimates of ultra-hot exoplanets are difficult to determine from photometric measurements only as these rely on assumptions regarding the planetary thermal emission~\citep[e.g.,][]{cowan_statistics_2011, schwartz_balancing_2015, bourrier_optical_2020}. Meanwhile the wide wavelength coverage of NIRISS has the benefit of simultaneously probing longer wavelengths dominated by the thermal flux of the planet. Nevertheless, the strict upper limit obtained on [Ti$/$H] (2$\sigma$ upper limit of $-$0.16 compared to [$\mathcal{V}/$H] = $1.24_{-0.35}^{+0.37}$ and [$\mathcal{R}/$H] = $1.17_{-0.19}^{+0.20}$) when the albedo is freely fit to the relatively high value of $A_g = 0.16_{-0.02}^{+0.02}$ is somewhat unexpected given that neutral Ti has been detected in the transmission spectrum of WASP-121b~\citep{prinoth_titanium_2025}. From an equilibrium chemistry perspective, the presence of atomic Ti in the atmosphere, even if underabundant, would suggest that TiO molecules can form as well~\citep{hoeijmakers_hot_2020}. While the argument could be made that the dayside is hotter than the terminator regions and hence TiO would be more dissociated, this is considered in the chemistry prescription of the retrieval. On the other hand, the TiO abundance is constrained (albeit to a low value) for the likely unphysical retrieval scenario of $A_g$ = 0. Possibly our model assuming a constant geometric albedo across the full NIRISS bandpass is too simplistic and the reality lies somewhere in between the $A_g$ = 0 and $A_g = 0.16_{-0.02}^{+0.02}$ scenarios. As a robustness check, we also tested retrievals subsequently masking out wavelengths lower than 0.65\,$\mu$m, 0.70\,$\mu$m, and 0.80\,$\mu$m to explore their impact on the inferred albedo.  For the three scenarios, we respectively recover $A_g = 0.15_{-0.03}^{+0.02}$, $A_g = 0.17_{-0.03}^{+0.03}$, and $A_g = 0.14_{-0.04}^{+0.04}$ indicating that the albedo inference is not driven by only the bluest data points. 

\begin{table}
\caption{Chemical equilibrium atmospheric retrievals results.} 
\vspace{-3mm}
\label{tab:retrieval_results}
\centering
\def\arraystretch{1.3}
\begin{tabular}{ccc}
\hline
Parameter & Chem.~($A_g$\,=\,0) & Chem.~($A_g$ fitted) \\
\hline
$[\mathcal{V}/$H$]$ & $1.11_{-0.28}^{+0.28}$ & $1.24_{-0.35}^{+0.37}$ \\
$[\mathcal{R}/$H$]$ & $1.31_{-0.16}^{+0.15}$ & $1.17_{-0.19}^{+0.20}$ \\
$[$Ti$/$H$]$ & $0.05_{-0.16}^{+0.15}$ & $<-$0.16 (2$\sigma$) \\
C$/$O & $0.48_{-0.16}^{+0.14}$ & $0.82_{-0.09}^{+0.05}$ \\
$A_g$ & $\cdots$ & $0.16_{-0.02}^{+0.02}$ \\
$A_\mathrm{HS}$ & $0.84_{-0.02}^{+0.02}$ & $0.89_{-0.03}^{+0.03}$ \\
$\log P_\mathrm{c}$\,(bar)& $>-$2.38\ (2$\sigma$) & $>-$2.17 (2$\sigma$) \\
Error inf.\ & $1.19_{-0.04}^{+0.04}$ & $1.16_{-0.04}^{+0.04}$ \\
\hline
$[$Ti$/$H$]$ $-$ $[\mathcal{R}/$H$]$ & $-1.25_{-0.12}^{+0.11}$ & $<-$1.43 (2$\sigma$) \\
$[\mathcal{V}/$H$]$ $-$ $[\mathcal{R}/$H$]$ & $-0.19_{-0.16}^{+0.16}$ & $0.07_{-0.21}^{+0.22}$ \\
\hline
\multicolumn{3}{l}{\small $\mathcal{V}$ = Volatiles, $\mathcal{R}$ = Refractories}\\
\vspace{-3mm}
\end{tabular}\\ 
\end{table}

One possible source of a higher than expected albedo could be partial coverage of highly reflective clouds~\citep[e.g.,][]{coulombe_highly_2025}. While ultra-hot Jupiter daysides should have low albedos due to being too hot to form clouds~\citep{parmentier_thermal_2018, gao_aerosol_2020, helling_cloud_2021}, WASP-121b may be be cold enough for condensates to be circulated from the nightside via a strong jet~\citep[e.g.,][]{seidel_detection_2023, seidel_vertical_2025, prinoth_titanium_2025}, temporarily survive on the western (morning) limb of the dayside~\citep[e.g.,][see their Figure 2]{roman_clouds_2021}. Interestingly, the ultra-hot Jupiter KELT-20b$/$MASCARA-2b~\citep[$1.753\pm0.036$\,$R_{\rm Jup}$, T$_\mathrm{eq}$ = 2250\,K,][]{lund_kelt-20b_2017, talens_mascara-2_2018}, which has similar properties as WASP-121b ($1.753\pm0.036$\,$R_{\rm Jup}$, T$_\mathrm{eq}$ = 2350\,K), has a relatively reflective dayside, with $A_g = 0.26 \pm 0.04$ measured from CHEOPS and TESS~\citep{singh_cheops_2024}. While it is possible that this regime of ultra-hot Jupiters is more reflective, future eclipse observations extending to even bluer wavelengths with e.g., HST$/$UVIS will be useful to break the TiO/reflected light degeneracy~\citep[e.g.,][]{scandariato_phase_2022, radica_constraining_2025}.

The constraints on the hot spot area factor also depend on the reflected light treatment. Intuitively, one might expect $A_{\mathrm{HS}}$ to be larger in the case of $A_g = 0$ for extra thermal flux to compensate from the lack of extra light reflected from the star.  However, we find $A_{\mathrm{HS}}$ to be higher when an albedo is included (Fig.~\ref{fig:chem_retrieval}, bottom right panel), to instead compensate for the overall hotter dayside retrieved, highlighting its correlation with the temperature structure. Overall our retrieved $A_{\mathrm{HS}}$ values slightly below unity could indicate that the emission emanating from the dayside is slightly concentrated near the substellar point rather than being uniform across the visible hemisphere. Such a picture would also be consistent with the large amplitude of the phase curve~\citep{mikal-evans_jwst_2023, morello_spitzer_2023}, which points to a strong temperature gradient around the planet. Interestingly, our retrieved hot spot area factor for WASP-121b is slightly less than the value measured for WASP-18b~\citep{coulombe_broadband_2023}, potentially owing to the lower gravity of WASP-121b ($\log g$ = 2.99) compared to WASP-18b ($\log g$ = 4.26) which would have reduced day-to-night heat redistribution~\citep{keating_uniformly_2019} and its infrared photosphere at deeper pressures where radiative timescales are longer likely making the atmosphere more uniform.

For the multiplicative error inflation factor, we recover a value of $1.16_{-0.04}^{+0.04}$ in our nominal retrieval including reflected light.  The factor being above unity could be due to remaining uncorrected systematics in the data, underestimated uncertainties, or our model being imperfect. One possible source of contamination for NIRISS SOSS data are background sources overlapping with the extracted spectral traces (Fig.~\ref{fig:reduction_steps}, third panel). To verify that these do not bias our results, we test running a retrieval (with $A_g$ fitted) where we mask out all identified wavelengths regions that have any overlap with background contaminants (Section~\ref{subsec:data_reduction}), finding nearly identical results (e.g., $[$$\mathcal{V}/$H$]$ = $1.22_{-0.34}^{+0.38}$, C$/$O = $0.80_{-0.12}^{+0.07}$, $A_g$ = $0.15_{-0.02}^{+0.02}$).

\subsection{The limitations of fitting a single global metallicity}\label{multi_metallicity}

\begin{figure}
    \centering
    \includegraphics[width=0.49\textwidth]{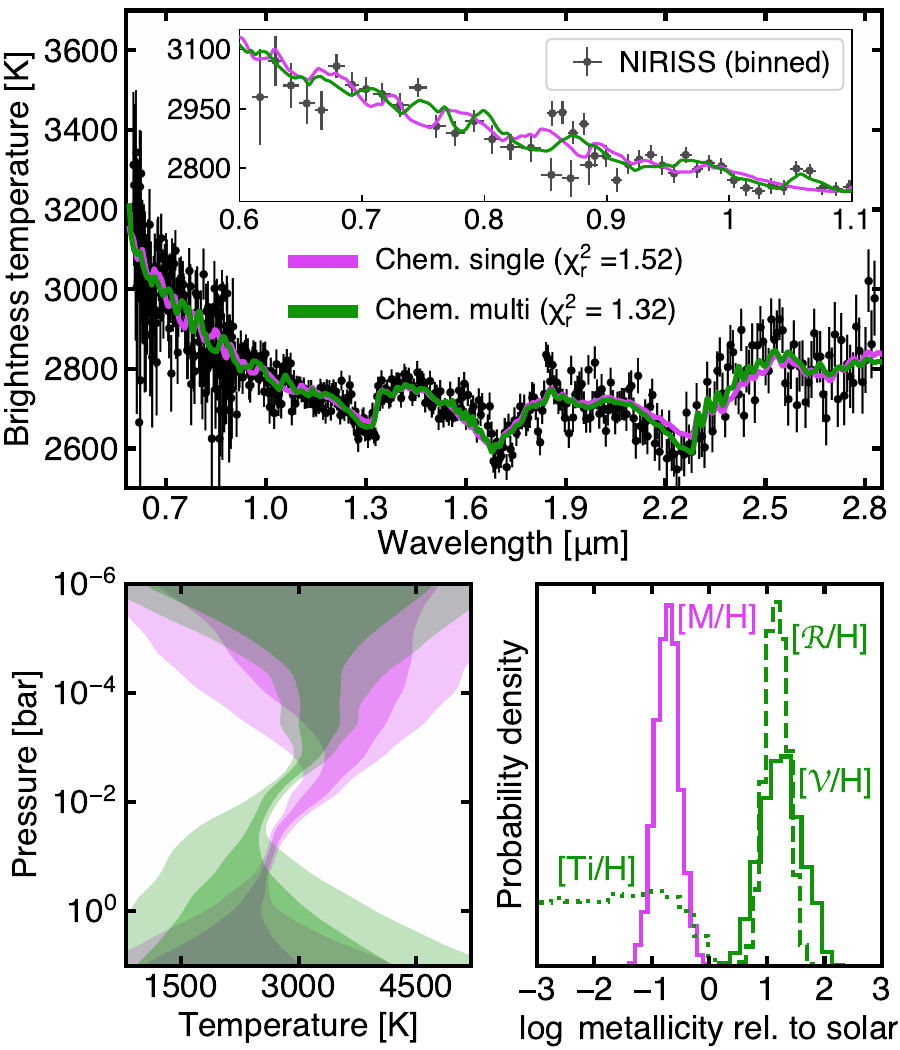}
    \vspace{-3mm}
    \caption{
    Importance of fitting more than a single global atmospheric metallicity in chemical equilibrium retrievals when probing spectral features affected by disequilibrium processes such as cold-trapping.   
    Top: Brightness temperature NIRISS spectrum (black points) overplotted with the best fit model from chemical equilibrium retrieval analyses only fitting for a single overall metallicity (pink), or fitting for volatiles, refractories, and titanium separately (green).  The inset panel shows a zoom in on the 0.6 to 1.1\,$\mu$m region, with the data now binned to $R=100$ for order 1 and $R=50$ for order 2 (grey points) for better visualisation.
    Bottom left: Retrieved TP profile in both cases (with $A_g$ fitted) showing that the single-metallicity case prefers an inversion much deeper in the atmosphere.
    Bottom right: Recovered abundances highlighting the importance of treating titanium as a separate parameter relative to other species in chemical equilibrium retrievals. Even when allowed to vary independently within a retrieval (``Chem.~multi'', green), the volatile ([$\mathcal{V}/$H], solid) and refractory ([$\mathcal{R}/$H], dashed) metallicities are recovered at similar enriched values while [Ti$/$H] (dotted) is significantly depleted in comparison. Enforcing all species to vary according to a single global metallicity (``Chem.~single'', [M$/$H], pink) and only allowing the C$/$O ratio to vary can result in some biased average of the depleted [Ti$/$H] and super-solar other metals to be inferred.
    }
    \label{fig:chem_multi}
    \vspace{-3mm}
\end{figure}

In this section we highlight the importance of fitting more than a single global metallicity in chemical equilibrium retrievals for atmospheres that may be affected by non-equilibrium processes such as cold-trapping.  For WASP-121b, it has long been suggested that TiO is cold-trapped on its colder nightside in order to explain numerous non detections~\citep{evans_optical_2018, hoeijmakers_hot_2020, hoeijmakers_mantis_2024, merritt_non-detection_2020, pelletier_crires_2025}. More recently, atomic Ti was detected in the transmission spectrum of WASP-121b, although its signature was significantly weaker than expected~\citep{prinoth_titanium_2025}. In either case, if some titanium is missing, this would make any chemical network overpredict the TiO abundance with respect to other species for a given metallicity, which could bias any inferred results.  This is because chemical networks typically scale elemental abundances uniformly, maintaining solar-like ratios unless specifically varied such as for carbon relative to oxygen via the C$/$O ratio. A similar problem can ensue if the volatile-to-refractory ratio in the planetary atmosphere differs from solar, for example as a result of its formation and accretion history~\citep{lothringer_new_2021, turrini_tracing_2021, chachan_breaking_2023}.

To investigate these potential pitfalls, we test running a retrieval fitting a single global metallicity as opposed to three separate parameters for volatiles, refractories, and titanium. We find that the single metallicity model, while initially appearing to be a decent fit, struggles to fully match the shape of the NIRISS spectrum, in particular the amplitude of the spectral features between 2.2 and 2.5\,$\mu$m (Fig.~\ref{fig:chem_multi}, top panel). In comparison, the multi metallicity model is significantly favoured (7.9$\sigma$) based on the calculated BPICS, despite the two extra fitted parameters. The difference in retrieved atmospheric properties is also drastic, with the global metallicity model instead inferring a metal-poor atmosphere with an inversion layer at much higher pressures (Fig.~\ref{fig:chem_multi}, bottom panels). This is mainly due to the equilibrium chemistry overpredicting the amount of TiO relative to other species, driving the metallicity lower to hide the resulting strong TiO bands under the H$^{-}$ continuum. This is somewhat analogous to the case of methane in warm ($\sim$700 - 800\,K) giant exoplanets, where assuming pure equilibrium chemistry and neglecting internal heating and vertical mixing can bias the retrieved temperature, which would otherwise mispredict the CO$/$CH$_4$ ratio~\citep{welbanks_high_2024, sing_warm_2024}. 

\begin{figure}
    \centering
    \includegraphics[width=0.49\textwidth]{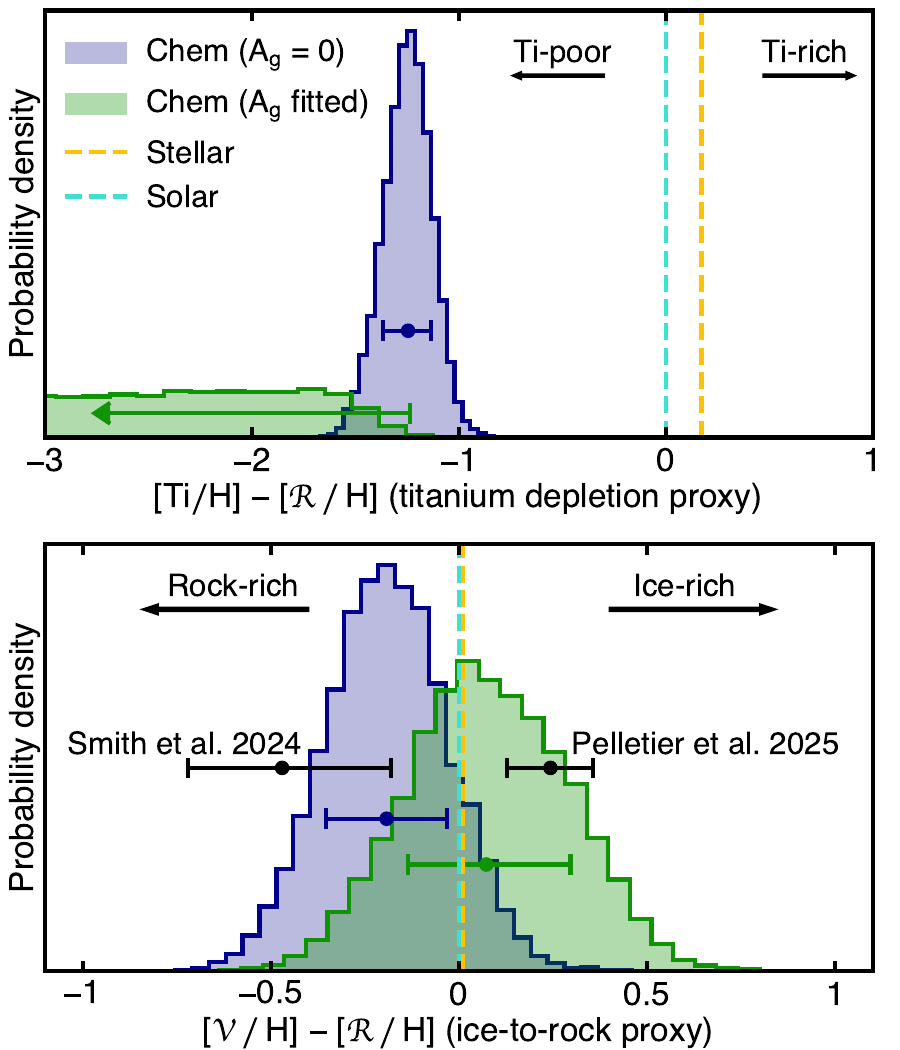}
    \vspace{-3mm}
    \caption{
    Relative abundance comparison for the chemical equilibrium retrievals of Fig.~\ref{fig:chem_retrieval} compared to the stellar~\citep[][dashed turquoise lines]{evans-soma_sio_2025} and solar~\citep[][dashed orange lines]{asplund_chemical_2021} values. Shown error bars are either 1\,$\sigma$ bounds or 3\,$\sigma$ upper limits.
    Top: Ratio of refractories compared to TiO (a proxy for the degree of titanium depletion). Compared to other refractory species such as VO included in [$\mathcal{R}/$H], TiO has a higher condensation temperature~\citep{lodders_solar_2003}, and hence its underabundance relative to other refractories may be indicative of some titanium being cold-trapped on the colder nightside of WASP-121b. 
    Bottom: Ratio of volatiles relative to refractories (a proxy for the ice-to-rock ratio). While our measurements are well consistent with a stellar/solar composition, we also cannot rule out the previous claims of WASP-121b harbouring a slightly rock-rich~\citep{lothringer_new_2021, smith_roasting_2024} or ice-rich~\citep{pelletier_crires_2025, evans-soma_sio_2025} atmosphere.
    }
    \label{fig:Ti_ice}
    \vspace{-3mm}
\end{figure}

\begin{figure*}[t!]
\begin{center}
\includegraphics[width=\linewidth]{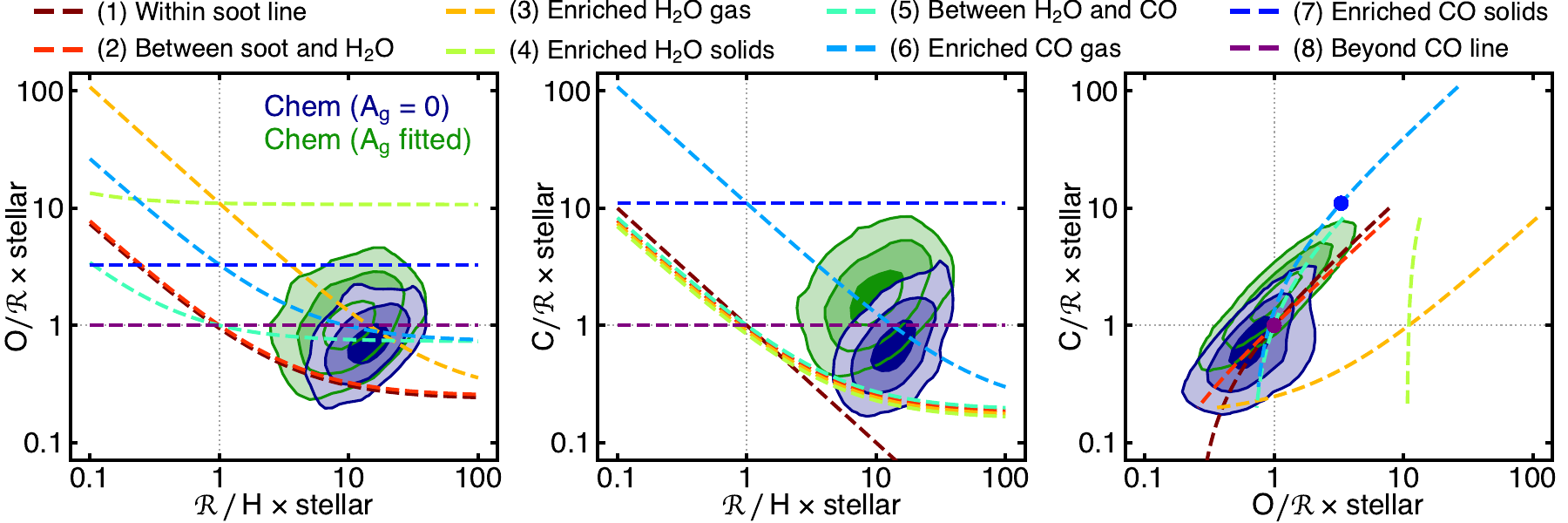}
\end{center}
\vspace{-4mm}
\caption{
Measured atmospheric composition of WASP-121b from the chemical equilibrium retrievals (blue and green 1, 2, and 3$\sigma$ contours) compared to predictions from the disc formation model of \cite{chachan_breaking_2023}. Here each dashed coloured line (or solid point in the case of a projection) corresponds to an envelope composition that WASP-121b would have accreted for different theoretical protoplanetary disc formation locations relative to the soot, H$_2$O, and CO snowline. Scenario 1 corresponds to the inner disc where all carbon is in the gas phase.  Scenario 2 is between the soot and H$_2$O snowlines where carbon can be held in solids.  Scenarios 3 and 4 are slightly within and exterior to the H$_2$O snowline where the gas and solid phases can, respectively, be enriched in H$_2$O. Scenario 5 is between the H$_2$O and CO snowlines where H$_2$O will be in ice form but CO remains in gas form. Scenarios 6 and 7 are inward and outward of the CO snowline where gas and solids are CO enriched. Scenario 8 is outwards of the CO snowline where even CO is condensed and accreted via solids. All abundance ratios are relative to the stellar values (dotted grey lines). While the scenarios most consistent with the measured atmospheric composition of WASP-121b (i.e., passing through the contours in all three panels) are formation in a CO-rich gas environment (scenario 6, light blue dashed line) or beyond the CO snowline (scenario 8, purple dashed line), most other scenarios from this model remain consistent at 3$\sigma$.
}
\label{fig:formation}
\vspace{-3mm}
\end{figure*}

 \subsubsection{Titanium cold-trapping}\label{subsubsec:titanium}
 
If allowed to vary separately, we find that [Ti$/$H] is strongly depleted relative to other species. Compared to the refractory metallicity parameter [$\mathcal{R}/$H] (mostly driven by VO), titanium is underabundant by
a factor of $26.1_{-5.9}^{+8.3}$ times stellar ($17.6_{-4.0}^{+5.6}$ times solar) in the case of assuming no reflected light, or is at least depleted by a factor of 25.6 times stellar (17.3 times solar) if $A_g$ is freely fit (3$\sigma$ limit).  Similar depletion levels are obtained if comparing [Ti$/$H] to [$\mathcal{V}/$H]. In both cases, TiO would seem to be depleted relative to other refractories by a factor of ten or more (Fig.~\ref{fig:Ti_ice}, top panel). Potentially this underabundance could explain previous non detections~\citep[e.g.,][]{evans_optical_2018, merritt_non-detection_2020, hoeijmakers_hot_2020, hoeijmakers_mantis_2024}, while still allowing for some gas-phase TiO to be present to help drive WASP-121b's stratosphere alongside VO, SiO, H$^{-}$, and numerous atomic metals and ions detected in its atmosphere~\citep{sing_hubble_2019, merritt_inventory_2021, lothringer_uv_2022, evans-soma_sio_2025, gapp_wasp-121_2025}. In tandem with atomic Ti being present (but also seemingly underabundant) in the transmission spectrum of WASP-121b~\citep{prinoth_titanium_2025}, it could be that the cold-trap mechanism that had been proposed to explain the TiO depletion is only partly efficient -- trapping a significant fraction of the Ti budget to the nightside and/or deeper atmospheric layers, but not its entirety. In this scenario, Ti species would condense out of the gas phase on the nightside and, depending on the vertical mixing strength, gravitationally settle to deeper atmospheric layers~\citep{spiegel_can_2009}.  Even when recirculated on the dayside, higher pressure regions of the atmosphere shielded from direct stellar irradiation may be cool enough for Ti to remain condensed, with efficient vertical mixing potentially allowing some of the Ti to sublimate back to the observable gas phase~\citep{parmentier_3d_2013}, explaining the apparent underabundance of both TiO and neutral Ti in observations~\citep[e.g.,][]{pelletier_crires_2025, prinoth_titanium_2025}. Although the dayside of WASP-121b is too hot for even highly refractory elements such as titanium to condense (see dotted line in Fig.~\ref{fig:chem_retrieval}, top left panel), this is not the case for the colder nightside~\citep{mikal-evans_diurnal_2022, evans-soma_sio_2025}. However, why TiO would condense but not VO or any of the other numerous refractory metals detected in the atmosphere of WASP-121b at high spectral resolution that do not show any sign of depletion~\citep[e.g., Mg, Fe, Cr][]{gibson_relative_2022, maguire_high-resolution_2023, gandhi_retrieval_2023} despite their difference of only $\sim$150 - 300\,K in condensation temperature~\citep{lodders_solar_2003} is interesting from a cloud formation perspective. While one might expect the initial condensation of TiO$_2$ grains to act as cloud condensation nuclei favouring the condensation of other slightly less refractory metals~\citep{fegley_atmospheric_1996, gao_aerosol_2020}, measurements seem to indicate that this is not the case on WASP-121b.

 \subsubsection{Volatile-to-refractory ratio}\label{subsubsec:ice_rock}

While the amount of titanium relative to other metals is strongly favoured to be lower if freely fit, the volatile [$\mathcal{V}/$H] and refractory [$\mathcal{R}/$H] metallicity parameters contrastingly both converge to similarly enriched values even though fitted independently (Table~\ref{tab:retrieval_results}). Interestingly, their relative proportions (a proxy for the ice-to-rock ratio), measured to be [$\mathcal{V}/$H] $-$ [$\mathcal{R}/$H] = $-0.19_{-0.16}^{+0.16}$ for $A_g = 0$ and [$\mathcal{V}/$H] $-$ [$\mathcal{R}/$H] = $0.07_{-0.21}^{+0.22}$ in the fitted albedo case, remains broadly consistent with the solar and stellar values (Fig.~\ref{fig:Ti_ice}, bottom panel). A similar degree of enrichment for both volatile and refractory metals while maintaining a stellar-like ratio was also recently inferred for the dayside atmosphere of WASP-121b at longer wavelengths using JWST$/$NIRSpec~\citep{evans-soma_sio_2025}. The slightly lower volatile-to-refractory ratio measured when not including a reflected starlight light component reflects the need for more refractory metals with strong optical opacity (e.g., VO) to be added to fit the excess higher brightness temperature measured at short wavelengths (Fig.~\ref{fig:chem_retrieval}, top right panel). Given the uncertainties of the measurements, however, we cannot rule out the slightly rock-rich~\citep{lothringer_new_2021, smith_roasting_2024} or ice-rich~\citep{pelletier_crires_2025, evans-soma_sio_2025} scenarios previously reported for WASP-121b. The lack of any significant deviation from the stellar volatile-to-refractory ratio could mean that WASP-121b would have accreted a similar proportion of ices and rocks over the course of its evolution~\citep{turrini_tracing_2021}.

While traditional chemical equilibrium retrievals fitting only for the C$/$O ratio and a single global metallicity as composition parameters can be an attractive option owing to the low number of free parameters, from our findings we caution against this practice for analysing data sets of atmospheres that may be affected by disequilibrium processes such as cold-trapping. Instead, elemental species in chemical equilibrium retrievals should ideally be fitted either in appropriate groups or entirely separately~\citep[e.g.,][]{smith_roasting_2024, gapp_wasp-121_2025, evans-soma_sio_2025, pelletier_crires_2025}.

\subsection{Link to planet formation}\label{formation}

In this section we make the assumption that the observed dayside atmospheric composition of WASP-121b is representative of the bulk envelope and explore scenarios in which this composition is a direct product of formation. This revolves around the concept that a planet's accretion history can leave an imprint onto its present-day atmospheric composition~\citep{mordasini_imprint_2016}. Overall we find that the constraints on carbon, oxygen, and refractories (via VO) provided by the NIRISS data for WASP-121b best match formation scenarios far out in the disc, near or beyond the CO snowline. 

In ultra-hot Jupiter atmospheres, the vapourisation of condensed compounds results in the full chemical inventory of a planetary envelope becoming accessible via the gas phase. In particular, measuring refractories in tandem with volatiles gives access to the ice-to-rock ratio of the material from which giant planets are formed~\citep{lothringer_new_2021, chachan_breaking_2023, smith_roasting_2024, pelletier_crires_2025}, information that is difficult to obtain even for the giants in our Solar System. This provides an added dimension to the C$/$O ratio and (volatile) metallicity typically used for inferring formation pathways which can lead to degenerate scenarios~\citep{madhusudhan_toward_2014, turrini_tracing_2021, schneider_how_2021, pacetti_chemical_2022}.

To explore potential accretion scenarios for WASP-121b that would give rise to the observed elemental abundances, we predict the envelope composition for various formation locations using the protoplanetary disc model of \cite{chachan_breaking_2023}, to which we refer the reader for a full description. In this model, the end envelope compositions for eight formation locations in the protoplanetary disc are predicted relative to snowlines for soots, H$_2$O, and CO.  Here a snowline refers to an orbital distance in the protoplanetary disc beyond which a given species will be condense out of the gas phase.  For example, H$_2$O will be gaseous inward of the H$_2$O snowline, but condensed as ice beyond it, changing the respective proportions of oxygen held in the accreted solids and gas.  
Local metal enrichment of the gas slightly interior to snowlines can also occur owing to the sublimation of icy pebbles drifting inwards from the outer disc~\citep{oberg_excess_2016, booth_chemical_2017}.  Similarly, the outwardly diffusion and re-condensation of this enriched gas can enhance the ice content of the solid phase in colder regions slightly exterior to snowlines~\citep{stevenson_rapid_1988}. Taking the CO snowline as an example, the gas slightly inward of it will be enriched in CO (higher $\mathcal{V}$/H) while the solids slightly outward of it will be enriched in CO ice (higher $\mathcal{V}/\mathcal{R}$). The factor by which the local gas and solids are enhanced is set to ten in the model~\citep{chachan_breaking_2023}.

Comparing our measured abundances (normalised to the stellar values) to these predictions, we find that the best matching scenarios are if WASP-121b formed far out in the disc, either slightly within the CO snowline where it could have accreted CO-enriched gas, or beyond it where it could have accreted all of its metals from solids maintaining their stellar proportions (Fig.~\ref{fig:formation}).  This would imply that WASP-121b migrated a few tens of astronomical units between the present day and when it formed~\citep{chachan_breaking_2023}. While it is uncertain what mechanism would have driven such inward migration, one possibility would be a migration path involving limited interactions with the protoplanetary disc preventing significant amounts of significant water rich (low C/O) planetesimals concentrated near the midplane~\citep{lothringer_new_2021}. Such as scenario could have involved WASP-121b being kicked to a high eccentricity orbit via either planet-planet scattering~\citep{rasio_dynamical_1996} or Kozai cycles~\citep{kozai_secular_1962, lidov_evolution_1962} followed by orbital circularisation via tidal dissipation~\citep{nagasawa_formation_2008}. The present-day highly misaligned orbit of WASP-121b~\citep{delrez_wasp-121_2016, bourrier_hot_2020} could also be a remnant of such a dynamically active past. 

We note that while our measured composition of WASP-121b's atmosphere agrees best with the model predictions of envelope accretion occurring slightly within or beyond the CO snowline, we also cannot rule out several other formation scenarios which remain consistent with the measured abundances at 3$\sigma$ (Fig.~\ref{fig:formation}).  Nevertheless, the simultaneous measurement of refractories combined with C and O helps rule out scenarios that would otherwise be degenerate from measurements of volatile species alone. Combining these NIRISS data with other instruments such as NIRPSpec will be helpful to obtain more precise constraints and better constrain WASP-121b's accretion history. In particular, measuring elemental species with different condensation temperatures (e.g., ultra-volatile nitrogen or moderately refractory sulfur) would provide even more avenues for disentangling different formation scenarios~\citep[e.g..,][]{cridland_connecting_2020, turrini_tracing_2021, pacetti_chemical_2022}.

\section{Conclusions}\label{sec:conclusion}

We analysed the thermal emission spectrum of the ultra-hot Jupiter WASP-121b obtained as part of a full phase curve including two eclipses observed with NIRISS.  We extracted eclipse spectra independently using the \texttt{NAMELESS} and \texttt{exoTEDRF} pipelines, finding good overall consistency between them.  

Throughout all our tests, we find that H$_2$O, CO, VO, and the bound-free contribution of H$^{-}$ are always necessary to fit the spectrum. Molecular bands are seen in emission, best matched by atmosphere models with temperature increasing from $\sim$2600\,K to upwards of 3100\,K at lower pressures. Our results confirm the previous detections of H$_2$O and CO made from both transmission and emission on WASP-121b~\citep{evans_detection_2016, evans_ultrahot_2017, mikal-evans_confirmation_2020, mikal-evans_diurnal_2022, wardenier_phase-resolving_2024, smith_roasting_2024, pelletier_crires_2025, evans-soma_sio_2025, gapp_wasp-121_2025, bazinet_quantifying_2025}. The detections of VO and H$^{-}$ (bound-free) but lack of strong TiO bands also gives new insights into which chemical species are the dominant optical absorbers, likely implying that TiO is not the principal driver of WASP-121b's stratosphere. However, TiO may still play a more important role if the mechanism driving its underabundance here is not as efficient, for example on a higher gravity~\citep[e.g., WASP-18b][]{coulombe_broadband_2023} or hotter~\citep[e.g., WASP-189b][]{prinoth_titanium_2022, prinoth_time-resolved_2023} giant planets. We stress that fitting TiO, VO, and H$_{\mathrm{bf}}^{-}$ simultaneously is critical~\citep[e.g.,][]{arcangeli_h-_2018, mikal-evans_emission_2019} as neglecting, for example, the bound-free component of H$^{-}$ can otherwise lead to unphysically high abundance inferences of other optical absorbers to compensate for missing opacity sources~\citep[e.g.,][]{evans_ultrahot_2017, changeat_five_2022, ouyang_detection_2023}. Most likely previous non-detections of VO from ground-based high resolution spectroscopy~\citep{merritt_non-detection_2020, hoeijmakers_hot_2020, hoeijmakers_mantis_2024} can be at least partially attributed to the imperfect VO line list at high resolution~\citep{regt_quantitative_2022, pelletier_vanadium_2023, maguire_high_2024}.

When assuming WASP-121b to be perfectly dark (non-reflective), we find that TiO is required to match the observed rise in brightness temperature at short wavelengths. If freely fitting for the albedo, however, the models favour reflected light over the combination of TiO and a slightly stronger thermal inversion to fit the spectrum, finding a geometric albedo of $A_g = 0.16_{-0.02}^{+0.02}$ and setting a strict upper limit on the TiO abundance. Regardless of whether reflected light is included, TiO is always found to be underabundant relative to other species by at least a factor of ten, although trace amounts of titanium need be present in the atmosphere to explain previous literature results~\citep{prinoth_titanium_2025}. Given its higher condensation temperature, the cause of titanium being underabundant could be the results of a partial cold-trap removing a large portion of the Ti budget from the gas phase. Ti-species being only present in trace amounts (10$\times$ depleted or more) would also naturally explain the numerous previous claimed non-detections of Ti and TiO~\citep{evans_optical_2018, hoeijmakers_hot_2020, hoeijmakers_mantis_2024, merritt_non-detection_2020, wilson_geminigmos_2021, gibson_relative_2022, maguire_high-resolution_2023, pelletier_crires_2025}. While our secondary eclipse observations cannot adequately disentangle its contribution from reflected light, the presence of TiO can be assessed through WASP-121b's NIRISS transmission spectrum (MacDonald et al.~in prep.).

Other than TiO, we find the atmosphere of WASP-121b to be metal rich, with both volatiles (e.g., H$_2$O and CO) and refractories (e.g., VO) being roughly equally enriched relative to solar/stellar by roughly a factor of ten. No statistically significant deviation is found between the volatile-to-refractory ratio of WASP-121b and that of the Sun or host star. These findings are notably irrespective of the albedo treatment, unlike our inferred C$/$O ratio which ranges from being consistent with stellar (C$/$O = $0.48_{-0.16}^{+0.14}$) if $A_g = 0$ to being super-stellar (C$/$O = $0.82_{-0.09}^{+0.05}$) if reflected light is included. Interestingly, the higher C/O ratio associated with a significantly non-zero albedo of $A_g = 0.16_{-0.02}^{+0.02}$ throughout the NIRISS bandpass is more in line with previous characterisations of the dayside atmosphere of WASP-121b finding super-stellar values of C$/$O between 0.59 and 0.87~\citep{changeat_is_2024}, C$/$O = $0.70_{-0.10}^{+0.07}$~\citep{smith_roasting_2024}, C$/$O = $0.73_{-0.08}^{+0.07}$~\citep{pelletier_crires_2025}, and C$/$O = $0.92_{-0.03}^{+0.02}$~\citep{evans-soma_sio_2025}. However, it remains unclear what would cause the dayside atmosphere of WASP-121b to have such a relatively elevated albedo given its high temperature preventing the formation of more reflective condensates on the dayside.

We find that the inclusion of H$_2$O dissociation in the models, although expected to be significant in the dayside atmosphere of WASP-121b, is not strictly necessary only favoured at the $\sim$2\,$\sigma$ level) to adequately fit the spectrum. However, neglecting it in free retrievals can result in significant biases in inferred temperature structure, continuum pressure level, and abundances. In particular, the relative proportions of H$_2$O and CO can still be adjusted to match the data, although the abundances would reflect the highly dissociated photospheric composition rather than that of the deep atmosphere.

For retrieval analyses enforcing chemical equilibrium, we find that only varying the overall metallicity and C$/$O ratio egregiously fails at fitting the NIRISS eclipse spectrum of WASP-121b. In particular, the underlying assumption that titanium is in solar proportions in the models vastly overestimates the TiO abundance relative to other species such as VO due to the chemical network being unable to replicate the non-equilibrium process of a nightside (partial) cold-trap. Instead, species affected by disequilibrium processes, such as Ti in this case, should be fitted as their own parameters in chemical equilibrium retrievals.  We find that neglecting this for WASP-121b results in orders of magnitude biases in inferred atmospheric properties.

Overall the NIRISS data provides a high quality view of the 0.6 to 2.8\,$\mu$m dayside spectrum of WASP-121b, producing precise constraints on the composition and thermal structure. Nevertheless, the inference of some atmospheric properties remains dependent on some model assumptions. Future joint analyses with other instruments such as HST$/$UVIS, JWST$/$NIRSpec and JWST$/$MIRI that increase the wavelength coverage, or with high resolution spectrographs such as IGRINS and CRIRES$^+$ to resolve spectral line shapes will ultimately help provide a more complete picture of the dayside atmosphere of WASP-121b.

\begin{acknowledgements}
The authors thank Thomas Evans-Soma for providing constructive feedback that helped us improve the quality of this manuscript.
S.P.\ thanks Vivien Parmentier (Observatoire de la C\^{o}te d'Azur) for useful discussions on titanium cold-trapping. This project was undertaken with the financial support of the Canadian Space Agency. The authors acknowledge support from the FRQNT through the Centre for Research in Astrophysics of Quebec. This project has been carried out within the framework of the National Centre of Competence in Research PlanetS supported by the Swiss National Science Foundation (SNSF) under grant 51NF40\_205606. The authors acknowledge the financial support of the SNSF.
R.J.M.\ acknowledges support by NASA through the NASA Hubble Fellowship grant HST-HF2-51513.001, awarded by the Space Telescope Science Institute, which is operated by the Association of Universities for Research in Astronomy, Inc., for NASA, under contract NAS 5-26555.
R.A.\ acknowledges the SNSF support under the Post-Doc Mobility grant P500PT\_222212 and the support of the Institut Trottier de Recherche sur les Exoplane\`etes (IREx).
L.D.\ acknowledges support from the Natural Sciences and Engineering Research Council (NSERC) and the Trottier Family Foundation.
D.J.\ is supported by NRC Canada and by an NSERC Discovery Grant.
C.P.-G acknowledges support from the E.~Margaret Burbidge Prize Postdoctoral Fellowship from the Brinson Foundation.
J.D.T.\ acknowledges funding support by the TESS Guest Investigator Program G06165.
This work is based on observations made with the NASA/ESA/CSA James Webb Space Telescope. The data were obtained from the Mikulski Archive for Space Telescopes at the Space Telescope Science Institute, which is operated by the Association of Universities for Research in Astronomy, Inc., under NASA contract NAS 5-03127 for JWST. These observations are associated with program GTO 1201.
This publication makes use of The Data \& Analysis Center for Exoplanets (DACE), which is a facility based at the University of Geneva (CH) dedicated to extrasolar planets data visualization, exchange and analysis. DACE is a platform of the Swiss NCCR PlanetS, federating the Swiss expertise in Exoplanet research. The DACE platform is available at \url{https://dace.unige.ch}.

\end{acknowledgements}

\bibliographystyle{aa}
\bibliography{W121_NIRISS}

\clearpage

\appendix

\onecolumn

\section{Secondary reduction analysis}\label{sec:jared_reduction}

\begin{figure*}[ht]
    \centering
    \includegraphics[width=\textwidth]{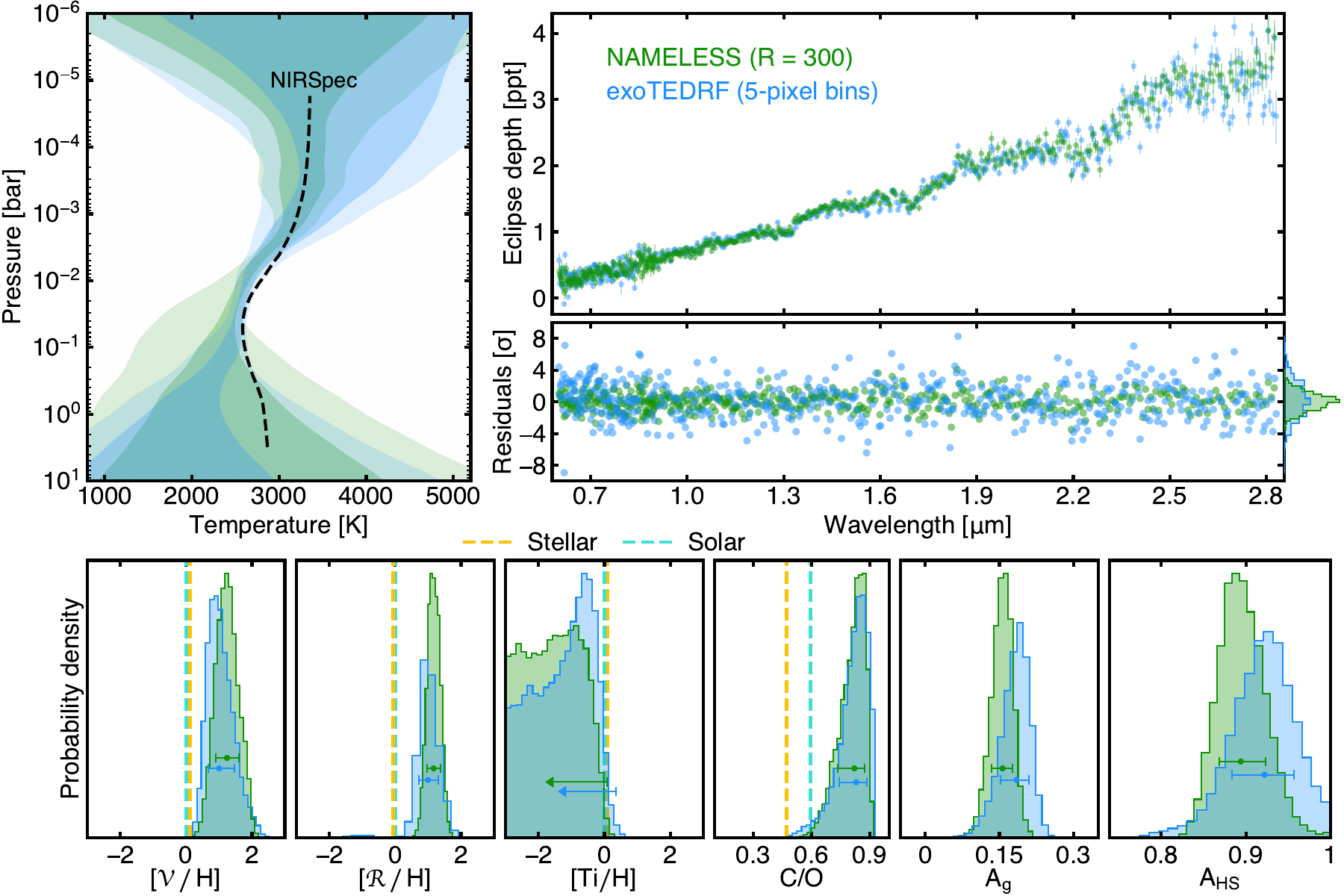}
    \caption{Similar as Fig.~\ref{fig:chem_retrieval}, but now comparing the chemical retrieval results with fitted albedo for the \texttt{NAMELESS} (green) and \texttt{exoTEDRF} (blue) pipelines.  The top right panels show the measured NIRISS eclipse depths and associated residuals when subtracting the respective best fit model of each retrieval. Both data reductions produce overall consistent results.
    }
    \label{fig:reduction_comparison}
\end{figure*}

While our main analysis used the NIRISS data reduction done with the \texttt{NAMELESS} pipeline, we also verified our results on a secondary data reduction using the \texttt{exoTEDRF} pipeline~\citep{feinstein_early_2023, radica_awesome_2023, radica_muted_2024, radica_exotedrf_2024}.  We refer the reader to \cite{splinter_precise_2025} for more details on the \texttt{exoTEDRF} reduction. The extracted eclipse depths and main retrieval results from both pipelines are compared in Fig.~\ref{fig:reduction_comparison}. While the retrieved parameter uncertainties are generally slightly smaller using \texttt{NAMELESS}, both reductions give similar constraints on all atmospheric parameters of WASP-121b. Given this consistency, we opt to focus on the results from the \texttt{NAMELESS} reduction in the main text.

\section{Individual eclipse analysis}\label{sec:individual_eclipses}

For our primary analysis, we extract the eclipse spectrum of WASP-121b by fitting the full observed phase curve. However, as the observed NIRISS phase curves contains two secondary eclipse, we can also extract spectra for these individually.  This can serve as a robustness test, to ensure that the spectra from each eclipse do not vary substantially.  Although WASP-121b has been suggested to be variable~\citep{wilson_geminigmos_2021, changeat_is_2024}, consecutive eclipses taken only one orbital period apart ($\sim$1.2 days) should still be within any atmospheric variability due to weather patterns that could induce changes at the percent level~\citep{tan_modelling_2024}, or up to 5 -- 10\%~\citep{changeat_is_2024} on a timescale of a several days.

\begin{figure*}[ht]
    \centering
    \includegraphics[width=\textwidth]{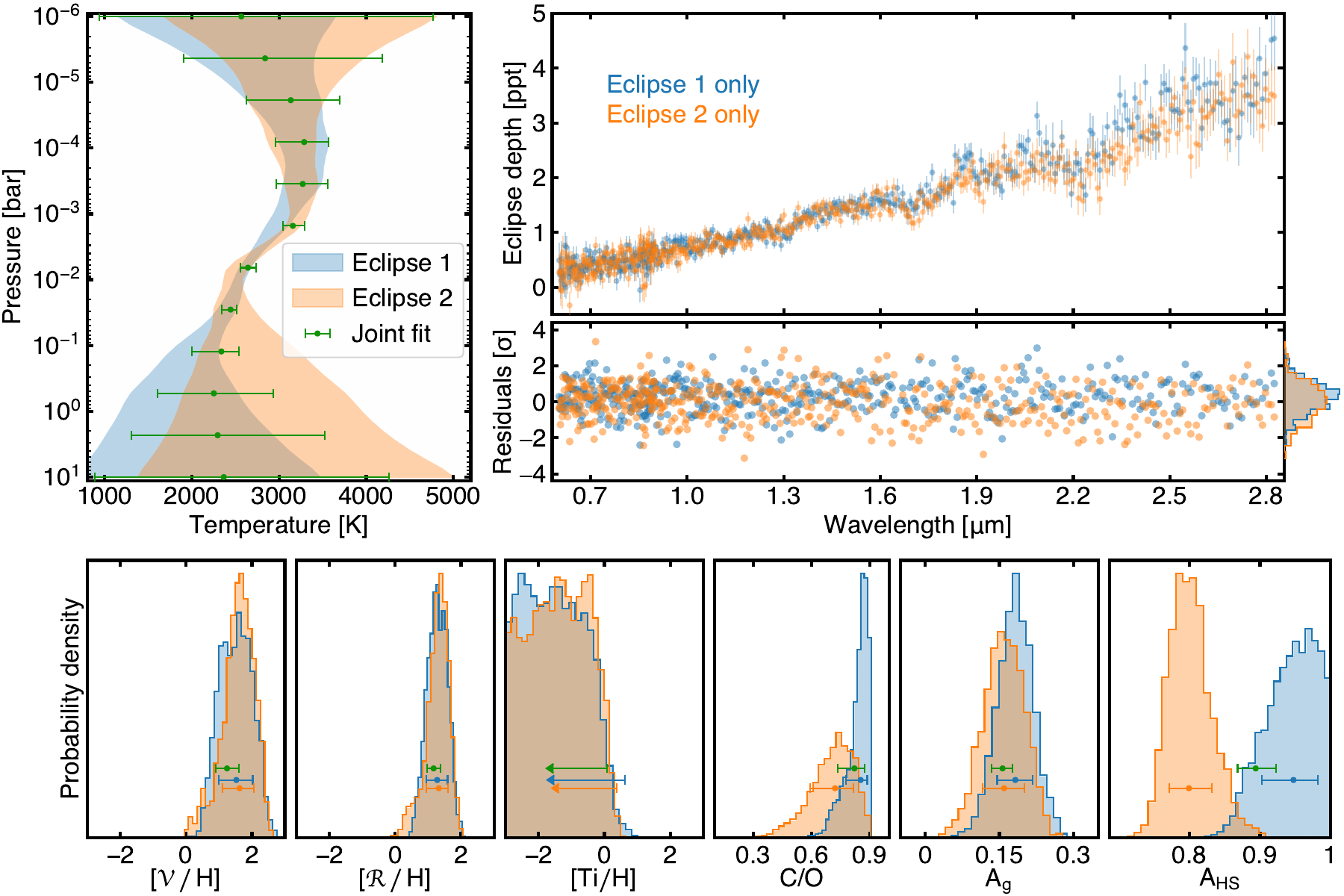}
    \caption{
    Similar as Fig.~\ref{fig:chem_retrieval}, but now comparing the the results from fitting eclipse spectra extracted from the first (blue) and second (orange) eclipses.  The retrieval setup is for our nominal case of assuming chemical equilibrium retrieval and fitting for the geometric albedo. Results are compared to the main retrieval results from fitting the full phase curve (green).
    Residuals are compared to the joint eclipse fit (Fig.~\ref{fig:eclipse_spec}). 
    With the exception of the hotspot area factor, similar atmospheric parameters are inferred for each eclipse separately.
    }
    \label{fig:eclipse_comparison}
\end{figure*}

We extract spectra for individual eclipses using the \texttt{NAMELESS} pipeline and following the same procedure as in Section~\ref{subsec:specLCF} for the case of a fixed resolving power $R$ = 300, but in each case only using data in and surrounding the eclipses ($\pm$1 transit duration).  
We find that while the first eclipse is slightly deeper at longer wavelengths, retrieved atmospheric properties are generally consistent (Fig.~\ref{fig:eclipse_comparison}). The largest difference is in the retrieved hotspot area factor $A_{\mathrm{HS}}$, which is smaller for the second eclipse, likely to account for the overall shallower eclipse depth. We note that the second eclipse contains the tilt event, which could be a source of bias.

\clearpage

\end{document}